\documentclass[%
 reprint,
superscriptaddress,
amsmath,amssymb,
aps,
pra,
rmp,
prstab,
prstper,
floatfix,
]{revtex4-2}

\usepackage{caption}
\usepackage{color}
\usepackage{graphicx}% Include figure files
\usepackage{dcolumn}% Align table columns on decimal point
\usepackage{bm}% bold math
\usepackage{subfigure}
\usepackage{float}
\begin{document}

\preprint{APS/123-QED}

\title{A Flexible GKP-State-Embedded Fault-Tolerant Quantum Computation Configuration Based on a Three-Dimensional Cluster State}

\author{Peilin Du}
\affiliation{
State Key Laboratory of Quantum Optics Technologies and Devices, Shanxi University, Taiyuan, 030006, China}%

\affiliation{
College of Physics and Electronic Engineering, Shanxi University,Taiyuan, 030006, China}

\author{Jing Zhang}
\email{zjj@sxu.edu.cn}
\affiliation{
State Key Laboratory of Quantum Optics Technologies and Devices, Shanxi University, Taiyuan, 030006, China%\\This line break forced% with \\
}%

\affiliation{
College of Physics and Electronic Engineering, Shanxi University,Taiyuan, 030006, China%\\This line break forced with \textbackslash\textbackslash
}%

\affiliation{
Collaborative Innovation Center of Extreme Optics, Shanxi University, Taiyuan, 030006, China%\\This line break forced with \textbackslash\textbackslash
}%

\author{Tiancai Zhang}

\affiliation{
State Key Laboratory of Quantum Optics Technologies and Devices, Shanxi University, Taiyuan, 030006, China%\\This line break forced% with \\
}%

\affiliation{
College of Physics and Electronic Engineering, Shanxi University,Taiyuan, 030006, China%\\This line break forced with \textbackslash\textbackslash
}%

\affiliation{
Collaborative Innovation Center of Extreme Optics, Shanxi University, Taiyuan, 030006, China%\\This line break forced with \textbackslash\textbackslash
}%

\author{Rongguo Yang}%

\affiliation{
State Key Laboratory of Quantum Optics Technologies and Devices, Shanxi University, Taiyuan, 030006, China%\\This line break forced% with \\
}%

\affiliation{
College of Physics and Electronic Engineering, Shanxi University,Taiyuan, 030006, China%\\This line break forced with \textbackslash\textbackslash
}%

\affiliation{
Collaborative Innovation Center of Extreme Optics, Shanxi University, Taiyuan, 030006, China%\\This line break forced with \textbackslash\textbackslash
}%

\author{Kui Liu}%

\affiliation{
State Key Laboratory of Quantum Optics Technologies and Devices, Shanxi University, Taiyuan, 030006, China%\\This line break forced% with \\
}%

\affiliation{
College of Physics and Electronic Engineering, Shanxi University,Taiyuan, 030006, China%\\This line break forced with \textbackslash\textbackslash
}%

\affiliation{
Collaborative Innovation Center of Extreme Optics, Shanxi University, Taiyuan, 030006, China%\\This line break forced with \textbackslash\textbackslash
}%

\author{Jiangrui Gao}%

\affiliation{
State Key Laboratory of Quantum Optics Technologies and Devices, Shanxi University, Taiyuan, 030006, China%\\This line break forced% with \\
}%

\affiliation{
College of Physics and Electronic Engineering, Shanxi University,Taiyuan, 030006, China%\\This line break forced with \textbackslash\textbackslash
}%

\affiliation{
Collaborative Innovation Center of Extreme Optics, Shanxi University, Taiyuan, 030006, China%\\This line break forced with \textbackslash\textbackslash
}%

\begin{abstract}
The integration of diverse quantum resources and the exploitation of more degrees of freedom provide key operational flexibility for universal fault-tolerant quantum computation. In this work, we propose a flexible Gottesman-Kitaev-Preskill-state-embedded fault-tolerant quantum computation architecture based on a three-dimensional cluster state constructed in polarization, frequency, and orbital angular momentum domains. Specifically, we design optical entanglement generators to produce three diverse entangled pairs, and subsequently construct a three-dimensional cluster state via a beam-splitter network with several time delays. Furthermore, we present a partially squeezed surface-GKP code to achieve fault-tolerant quantum computation and ultimately find the optimal choice of implementing the squeezing gate to give the best fault-tolerant performance (the fault-tolerant squeezing threshold is 11.5 dB). Our scheme is flexible, scalable, and experimentally feasible, providing versatile options for future optical fault-tolerant quantum computation architecture.
\end{abstract}

\maketitle

%\tableofcontents

\section{\label{sec:introduction}introduction}

Continuous-variable (CV) measurement-based quantum computation (MBQC) is a promising candidate for practical, scalable, universal, and fault-tolerant quantum computation, which relies on the deterministic generation of large-scale cluster states\cite{2025Scaling}. Since preparing three-dimensional (3D) cluster states is not only sufficient for universal MBQC but also indispensable for fault-tolerant quantum computation (FTQC)\cite{universalityPRL,Furusawa2024review}, numerous proposals have been put forward for their generation\cite{Fukui3D2020,Oliver3D2021,wu(2020),Du(2022)}.

However, only finite squeezing can be achieved in labs, leading to the inevitable introduction of Gaussian noise into quantum information during CV MBQC based on large-scale cluster states\cite{Larsen(2020)noiseanalysis,alexander2014noise,du2025complete}. Therefore, quantum error correction (QEC) based on the Gottesman-Kitaev-Preskill (GKP) code becomes indispensable, which first requires the preparation of GKP states\cite{Gottesman(2000)GKPcoding,GKPPRL}. Although numerous studies have proposed methods for generating GKP states\cite{Bour2021blue,KyungjooLow-Overhead,DahanGENGKP,AndersenGENGKP,UlrikGENGKP}, the practical experimental preparation of high-quality GKP states proves challenging\cite{FurusawaGKP2025,larsenGKP2025}. Several approaches have been explored to generate GKP states and inject them into cluster states\cite{Larsen(2021)PRXQuantum,Bour2021blue,ArbitraryPRL}. One feasible approach involves generating GKP states via Gaussian boson sampling, followed by injecting the generated GKP states into cluster states using optical switches\cite{Larsen(2021)PRXQuantum,Bour2021blue}. Another potential scheme is to generate GKP dumbbell states through Gaussian boson sampling and subsequently construct macronode cluster states using these GKP dumbbell states\cite{ArbitraryPRL}. Nevertheless, these methods generate GKP states probabilistically, thus requiring multiplexing for scheme scalability, and optical switches can introduce additional noise\cite{Switching-free}.  

Furthermore, the GKP code can only correct small displacement errors in phase space, with large displacement errors converted into Pauli errors in encoded GKP qubits. Residual Pauli errors can be further treated by combining the GKP code with qubit-level QEC codes, such as the surface code\cite{Larsen(2021)PRXQuantum,Fukui(2017)SURFACECODE,Noh(2019)SURFACE} and the Raussendorf-Harrington-Goyal (RHG) code\cite{Bour2021blue,TzitrinPRXQuantum(2021)}. The surface code is one of the most widely adopted qubit-level QEC codes due to its low fault-tolerant squeezing threshold and high tolerance to local errors\cite{SurfacecodesPRA}. Consequently, many FTQC schemes based on 3D cluster states and surface-GKP codes have been proposed\cite{Fukui(2017)SURFACECODE,Larsen(2021)PRXQuantum,biassurface,XZZXsurfacecode,KyungjooLow-Overhead}. For instance, an FTQC scheme based on a probabilistically generated complex 3D cluster state composed of GKP qubits was proposed, achieving a squeezing threshold of 9.8 dB by applying analog quantum error correction and the surface-GKP code\cite{Fukui(2017)SURFACECODE}. Recently, another FTQC scheme based on a 3D temporal cluster state was proposed, in which the surface-4-GKP code is employed to perform four GKP error corrections during each round of stabilizer measurements, yielding a squeezing threshold of 12.7 dB\cite{Larsen(2021)PRXQuantum}. Additionally, introducing bias into the surface-GKP code can further enhance its fault-tolerant performance, which requires the initial qubits to be biased GKP states\cite{biassurface,XZZXsurfacecode,KyungjooLow-Overhead}.

In this work, we demonstrate the generation of large-scale 3D cluster states involving the frequency, polarization, and orbital angular momentum degrees of freedom, whose elements can be flexibly selected from the EPR, hybrid, and GKP pairs. Thus, GKP states are embedded in the cluster states and are ready for use when needed, and our scheme circumvents the noise arising from the use of optical switches. Moreover, we propose the partially squeezed surface-GKP code to effectively reduce the fault-tolerant threshold by implementing a squeezing gate in a specific stabilizer measurement step, which is more convenient than preparing biased GKP states. The paper is organized as follows: Sec. II introduces the design of the optical entanglement generator and the construction of the large-scale 3D hybrid cluster states. Sec. III provides a detailed demonstration of fault-tolerant performance using our proposed partially squeezed surface-GKP code. Finally, Sec. IV summarizes the main results.

\section{Generation of large-scale 3D cluster states embedded with GKP states}

In this section, a flexible scheme for generating large-scale 3D cluster states embedded with GKP states is proposed. First, an optical entanglement generator (OEG) is designed to produce diverse types of entangled pairs. Subsequently, large-scale 3D cluster states can be constructed by adopting two such OEGs in beam splitter networks with time delays. The scheme is flexible because different 3D entanglement structures can be achieved by adopting different beam splitter combinations.  
\begin{figure}[htbp]
\centering\includegraphics[width=1\linewidth]{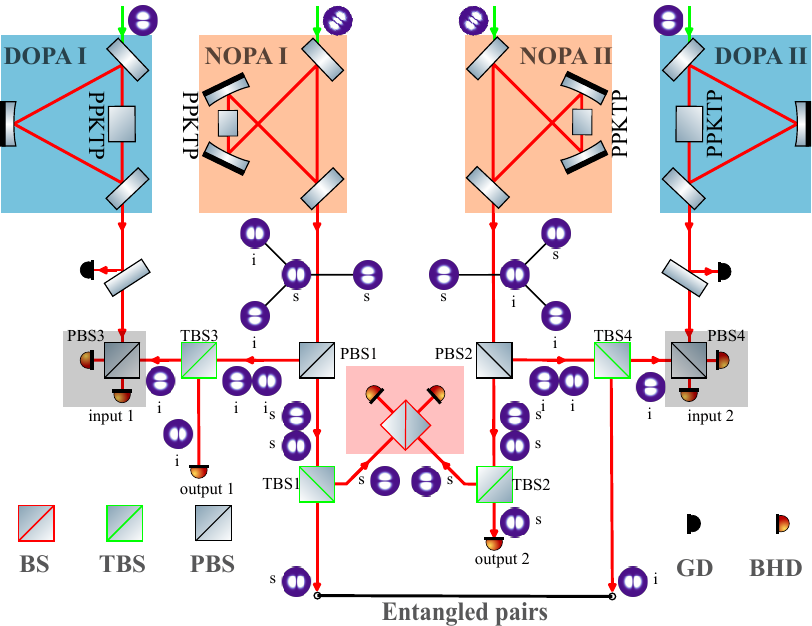}
\caption{The schematic of the optical entanglement generator. Symbols $s$ and $i$ denotes the signal and idler modes, respectively. GD: general-dyne detection; BHD: balanced homodyne detection.}
\label{fig1}
\end{figure}

\subsection{Generation of various entangled pairs}

The architecture of the OEG we developed is depicted in Fig.\ref{fig1}. This system is primarily composed of two non-degenerate optical parametric amplifiers (NOPAs) employed to generate cluster states, two degenerate optical parametric amplifiers (DOPAs) dedicated to preparing GKP states, and several optical components, including balanced beam splitters (BSs), polarization beam splitters (PBSs), and transverse-mode beam splitters (TBSs). 

As shown in Fig.\ref{fig1}, the NOPAs are bow-tie cavities containing type-II phase-matching nonlinear crystals (with the second-order nonlinear coefficient $\xi$). In these cavities, pump beams with frequency $2\omega_{0}$ are down-converted into signal and idler beams with frequencies $\omega_{0}\pm j \Delta$ and $\omega_{0}\mp j \Delta$, respectively. $\Delta$ represents the free spectral range of the NOPAs, and $j=0,1,2,... $. The interaction Hamiltonian under the undepleted pump approximation can be expressed as: 
\begin{equation}
\mathrm{H}=i\hbar\xi\sum_{j} \mathrm{G}\left(a^{\dagger}_{\left(\omega_0\pm j\Delta,s\right)}a^{\dagger}_{\left(\omega_0\mp j\Delta,i\right)}+\mathrm{H.c.}\right),
\end{equation}
where $a^{\dagger}_{\left(\omega_0\pm j\Delta,s\right)}$ and $a^{\dagger}_{\left(\omega_0\mp j\Delta,i\right)}$ are the creation operators of down-converted signal and idler beams with frequencies $\omega_0\pm j\Delta$ and $\omega_0\mp j\Delta$, respectively. Meanwhile, the pump beams of the NOPAs are spatially structured $\mathrm{HG}_{02}^{45^{\circ}}$ beams, which can be expressed in the Hermite-Gaussian basis as:
\begin{equation}
\mathrm{HG}_{02}^{45^{\circ}}=\frac{1}{2} \mathrm{HG}_{02}+\frac{1}{\sqrt{2}} \mathrm{HG}_{11}+\frac{1}{2} \mathrm{HG}_{20}.
\end{equation}
According to the selection rules\cite{rulesoftransmode}, the following four nonlinear processes are allowed:
\begin{eqnarray}
\begin{aligned}
    &\mathrm{HG}^{b}_{02}\left(2\omega_{0}\right)\to \mathrm{HG}^{s}_{01}\left(\omega_{0}\pm j\Delta\right)+\mathrm{HG}^{i}_{01}\left(\omega_{0}\mp j\Delta\right),\\ 
  &\mathrm{HG}^{b}_{11}\left(2\omega_{0}\right)\to \mathrm{HG}^{s}_{01}\left(\omega_{0}\pm j\Delta\right)+\mathrm{HG}^{i}_{10}\left(\omega_{0}\mp j\Delta\right),\\ 
   &\mathrm{HG}^{b}_{11}\left(2\omega_{0}\right)\to \mathrm{HG}^{s}_{10}\left(\omega_{0}\pm j\Delta\right)+\mathrm{HG}^{i}_{01}\left(\omega_{0}\mp j\Delta\right),\\ 
   &\mathrm{HG}^{b}_{20}\left(2\omega_{0}\right)\to \mathrm{HG}^{s}_{10}\left(\omega_{0}\pm j\Delta\right)+\mathrm{HG}^{i}_{10}\left(\omega_{0}\mp j\Delta\right).
\end{aligned}
\end{eqnarray}
Note that the down-converted signal and idler modes feature orthogonal polarizations, and the nonlinear processes obey the conservation laws of energy, momentum, and angular momentum. During these nonlinear processes in the NOPAs,  quadripartite GHZ states composed of the $\mathrm{HG}^{s}_{01}\left(\omega_{0}\pm j\Delta\right)$, $\mathrm{HG}^{i}_{01}\left(\omega_{0}\mp j\Delta\right)$, $\mathrm{HG}^{s}_{10}\left(\omega_{0}\pm j\Delta\right)$, and $\mathrm{HG}^{i}_{10}\left(\omega_{0}\mp j\Delta\right)$ modes can be generated (as demonstrated in one of our previous works \cite{liu2016GHZstate}). Such GHZ states can further be converted into star-type cluster states by transferring the amplitude and phase quadrature measurement (equivalent to Fourier transforms)\cite{Menicucci2011star}. In our scheme, these star-type states are adopted as the generated CV cluster states from the NOPAs.  

The DOPAs in Fig.\ref{fig1} are triangular cavities containing type-II phase-matching nonlinear crystals (with the second order nonlinear coefficient $\kappa$). The harmonic entanglement can be generated through the DOPA processes\cite{GrossePRL96HarmonicEntanglement,GrossePRLHarmonicEntanglement,guoAPL2012HarmonicEntanglement}, which enables the deterministic generation of GKP states by measuring the down-converted quadratures\cite{YanagimotoGKP(2023)PRXQuantum}. A spatiotemporal pump mode $\hat{b}=\mathrm{HG}_{01}\left(\omega_{0}\mp j\Delta\right)$, corresponding to the output idler modes from NOPAs after TBS3, can be down-converted into degenerate modes $\hat{a}=\mathrm{HG}_{01}\left((\omega_{0}\mp j\Delta)/2\right)$. In the case of frequency detuning, the Hamiltonian in the rotating frame can be written as: 
\begin{equation}
\mathrm{H}=i\hbar\kappa\sum_{j}\left(\hat{b}\hat{a}^{\dagger2}+\hat{b}^{\dagger}\hat{a}^{2}\right)+\delta\hat{a}^{\dagger}\hat{a},
\end{equation}
where $\delta$ is the frequency detuning, $\hat{b}$ ($\hat{b}^{\dagger}$) and $\hat{a}$ ($\hat{a}^{\dagger}$) are the annihilation (creation) operators of the pump and down-converted modes. By performing general-dyne measurements on the down-converted modes and providing feedback to the pump modes, GKP states can be deterministically generated. 
\begin{figure}[htbp]
\centering
\begin{subfigure}{
\centering
\includegraphics[width=\linewidth]{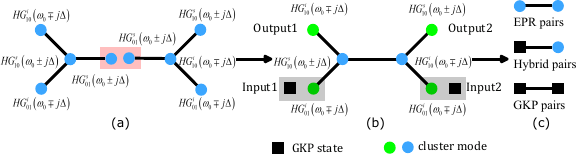}}
\end{subfigure}
\caption{Illustration of the entanglement structure evolution via the optical entanglement generator shown in Fig.1. The pink rectangular area represents the joint measurement for generating hexapartite cluster states, and the gray rectangular area represents the joint measurement for inputting GKP states. Solid circles represent the cluster modes, and black squares represent the GKP states.}
\label{Fig2}
\end{figure}

In Fig.\ref{fig1}, the pump beams $\mathrm{HG}_{02}^{45^{\circ}}$ pass through NOPA1 and NOPA2 and generate quadripartite star-type cluster states. Each star-type state comprises four modes: $\mathrm{HG}^{s}_{01}\left(\omega_{0}\pm j\Delta\right)$, $\mathrm{HG}^{i}_{01}\left(\omega_{0}\mp j\Delta\right)$, $\mathrm{HG}^{s}_{10}\left(\omega_{0}\pm j\Delta\right)$, and $\mathrm{HG}^{i}_{10}\left(\omega_{0}\mp j\Delta\right)$ modes, corresponding to Fig.\ref{Fig2}(a). Then, the signal and idler modes within these cluster states are separated via PBS1 and PBS2 due to their orthogonal polarizations. Subsequently, the $\mathrm{HG}_{01}$ and $\mathrm{HG}_{10}$ signal (idler) modes are further separated by TBS1(3) and TBS2(4), respectively. The separated $\mathrm{HG}_{01}$ signal modes are subjected to joint measurement, as denoted by the pink rectangular area in Fig.\ref{Fig2}(a). The separated $\mathrm{HG}_{01}$ idler modes from NOPA1(2) and the generated GKP state (corresponding to the black square in Fig.\ref{Fig2}(b)) from DOPA1(2) enter PBS3(4) and undergo homodyne detection, for inputting GKP states when error correction is needed, as denoted by the gray rectangular area. Finally, hexapartite cluster states are generated, as shown in Fig.\ref{Fig2}(b), which can be characterized as two entangled modes (blue circles, for quantum computation) connected to four branch modes (green circles, with two lower-branch modes for introducing GKP states to perform error correction and two upper-branch modes for outputting final results). 

Fig.\ref{Fig2} presents a detailed illustration of the entanglement structure evolution corresponding to Fig.\ref{fig1}. As depicted in Fig.\ref{Fig2}(a), the two star-type cluster states evolve into a hexapartite cluster state after a joint measurement. Notably, a series of hexapartite cluster states with different $j$ values can be generated in parallel within the frequency domain. Subsequently, the color of the down-converted modes used for inputting GKP states and outputting result states changes from blue to green, as shown in Fig.\ref{Fig2}(b). When the result states are not required, the upper-branch modes can be erased without disturbing other entangled modes by selecting the quadrature-amplitude measurement bases. Depending on the choice of measurement bases for the two lower-branch modes, three distinct entangled pairs can be obtained: identical bases produce EPR pairs, distinct bases generate GKP pairs, and a combination of one identical and one distinct basis yields hybrid pairs, as depicted in Fig.\ref{Fig2}(c). The EPR pairs, GKP pairs, and hybrid pairs generated herein are convenient for performing quantum computation, full teleportation-based GKP error correction, and partial teleportation-based GKP error correction, respectively\cite{BaragiolaStreamlined(2021)}. Therefore, by flexibly selecting measurement bases for branch modes, suitable large-scale GKP-embedded cluster states can be built to meet different requirements of quantum computation.

\subsection{Construction of large-scale 3D hybrid cluster states}
\begin{figure*}[ht]
\centering\includegraphics[width=1\linewidth]{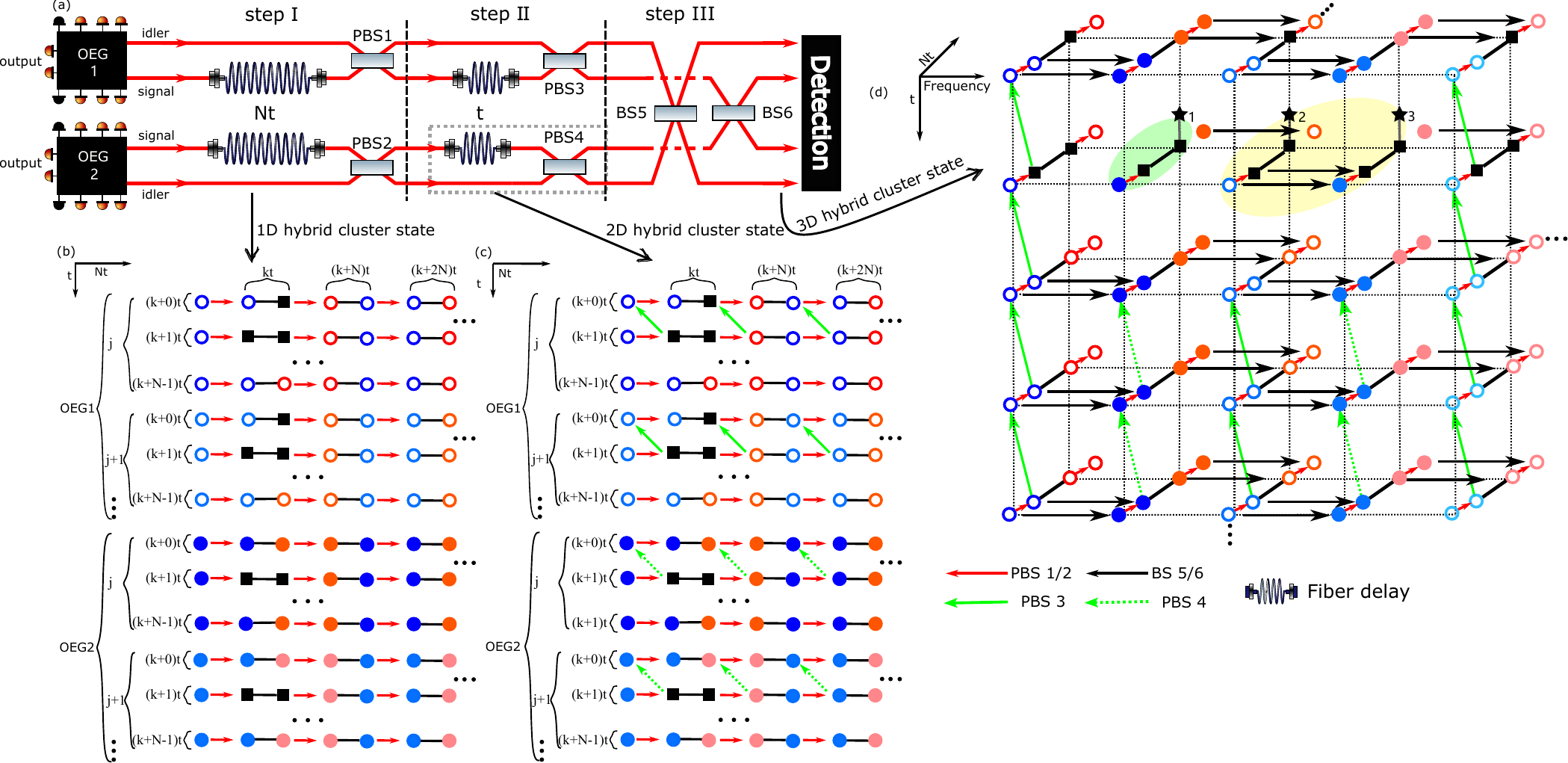}
\caption{The 3D cluster state construction. (a) Schematic diagram of our scheme. (b)–(d) Entanglement structures of 1D, 2D, and 3D cluster states in steps I–III, respectively. Red arrows: PBS1/PBS2; black arrows: BS5/BS6; green solid arrows: PBS3; green dotted arrows: PBS4. Filled (hollow) circles: modes from OEG1(OEG2). Black squares: GKP states; black stars: output ports.}
\label{Fig3}
\end{figure*}

Now we focus on the system depicted in Fig.\ref{Fig3}(a), which integrates two aforementioned OEGs with several time delays and is capable of generating large-scale 1D, 2D and 3D hybrid cluster states in sequence, corresponding to the structures depicted in Fig.\ref{Fig3}(b), Fig.\ref{Fig3}(c), and Fig.\ref{Fig3}(d), respectively. In Fig.\ref{Fig3}(a), two OEGs with NOPA pump frequencies $2\omega_{0}+\Delta$ and $2\omega_{0}-\Delta$ generate two sets of entangled pairs with flexible choices. The signal and idler beams of the OEGs are divided into time bins of a period $t$, where the bandwidths of the NOPA and DOPA cavities are much larger than $1/t$. Thus, a series of entangled pairs separated by a time interval $t$ can be deterministically created, with the entangled pairs in each time bin being independent of each other. As shown in Fig.\ref{Fig3}(a), the 1D, 2D, and 3D cluster states can be generated in steps I, II, and III, respectively.

In step I, the signal beams are delayed $Nt$, and the staggered signal and idler fields are coupled via PBS1 and PBS2 to generate $N$ sets of 1D dual-rail cluster states for each $j$ value, as shown in Fig.\ref{Fig3}(b). As an example, proper measurement bases are chosen for OEG1 to generate hybrid and GKP pairs at $kt$ and $(k+1)t$ time, respectively. For OEG2, EPR and GKP pairs are generated at $kt$ and $(k+1)t$ time, respectively. In step II, the signal beams are further delayed $t$, and the generated $N$ sets of staggered 1D dual-rail cluster states are correlated by PBS3 and PBS4, thereby forming 2D double-bilayer square-lattice cluster states, as shown in Fig.\ref{Fig3}(c). Finally, in step III, all 2D cluster states are further connected via BS5 and BS6 to form a 3D cluster state, as shown in Fig.\ref{Fig3}(d).

Specifically, the 3D cluster state shown in Fig.\ref{Fig3}(d) includes double-bilayer square-lattice and quadrail lattice structures in vertical and horizontal directions, respectively. These two entanglement structures are universal resources for CV MBQC\cite{Equivalentcluster}. It is evident in Fig.\ref{Fig3}(d) that three types of entangled pairs are present in the 3D entanglement structure. This means that hybrid and GKP pairs can be flexibly introduced to the 3D cluster state when error correction is needed by choosing appropriate measurement bases at the proper time. Moreover, one can output the quantum computation results at any time at the output ports. For instance, a single-mode (two-mode) Clifford gate and GKP error correction can be performed simultaneously\cite{BaragiolaStreamlined(2021)}, and the results can be obtained at the output ports 1 (2 and 3), as shown by the light green (light yellow) area in Fig.\ref{Fig3}(d). 

\begin{figure}[ht!]
\centering\includegraphics[height=0.15\textheight,width=0.6\linewidth]{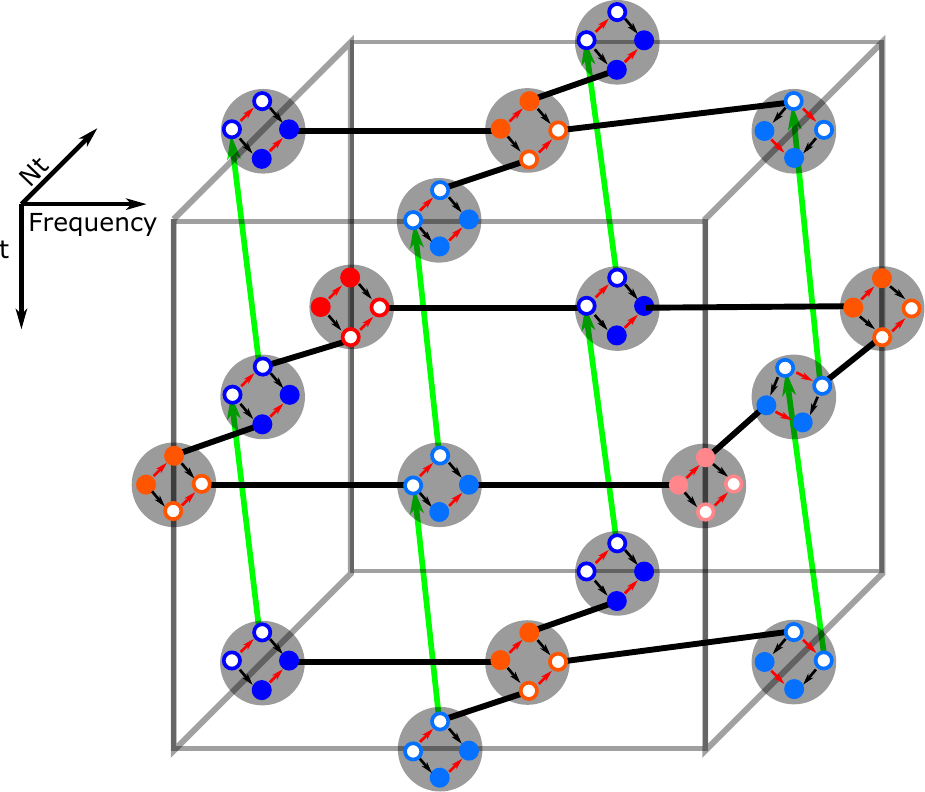}
\caption{Another 3D cluster state structure generated by our scheme -- the 3D macronode RHG cluster state. The gray circle areas represent the macronode modes and all colored arrows (denoting BSs) and circles (denoting output modes from OEGs) are identical to those in Fig.3.}
\label{Fig4}
\end{figure}

In addition, if the time-delay and PBS4 (framed by the gray dotted box in Fig.\ref{Fig3}(a)) are removed in step II, a 3D macronode RHG cluster state can be generated after step III, as shown in Fig.\ref{Fig4}. Each macronode consists of four modes coupled by the four PBSs and correlates with the other four macronodes, forming a quadrail lattice structure. By measuring three modes of each macronode in the quadrature amplitude basis to decouple them from the central mode\cite{TzitrinPRXQuantum(2021)}, the 3D macronode RHG cluster state can be reduced to a 3D canonical RHG cluster state, enabling topologically protected quantum computation. When the central mode of each macronode is chosen to be the GKP state, FTQC can also be achieved.

\section{Fault-tolerant quantum computation using the partially squeezed surface-GKP code}

The universal quantum computation and GKP error correction can be implemented simultaneously based on the generated 3D cluster state\cite{teleportationQEC,BaragiolaStreamlined(2021)}. However, the GKP error correction can only correct small displacement errors (less than $\sqrt{\pi}/2$) in phase space, while large displacement errors are converted into Pauli errors of data GKP qubits and can be further treated by combining GKP code with certain qubit error correction codes towards FTQC. Here, we present the partially squeezed surface-GKP code to achieve FTQC and analyze the fault-tolerant performance of introducing squeezing operations at different steps during the syndrome measurement. 
\begin{figure}[]
\centering\includegraphics[width=\linewidth]{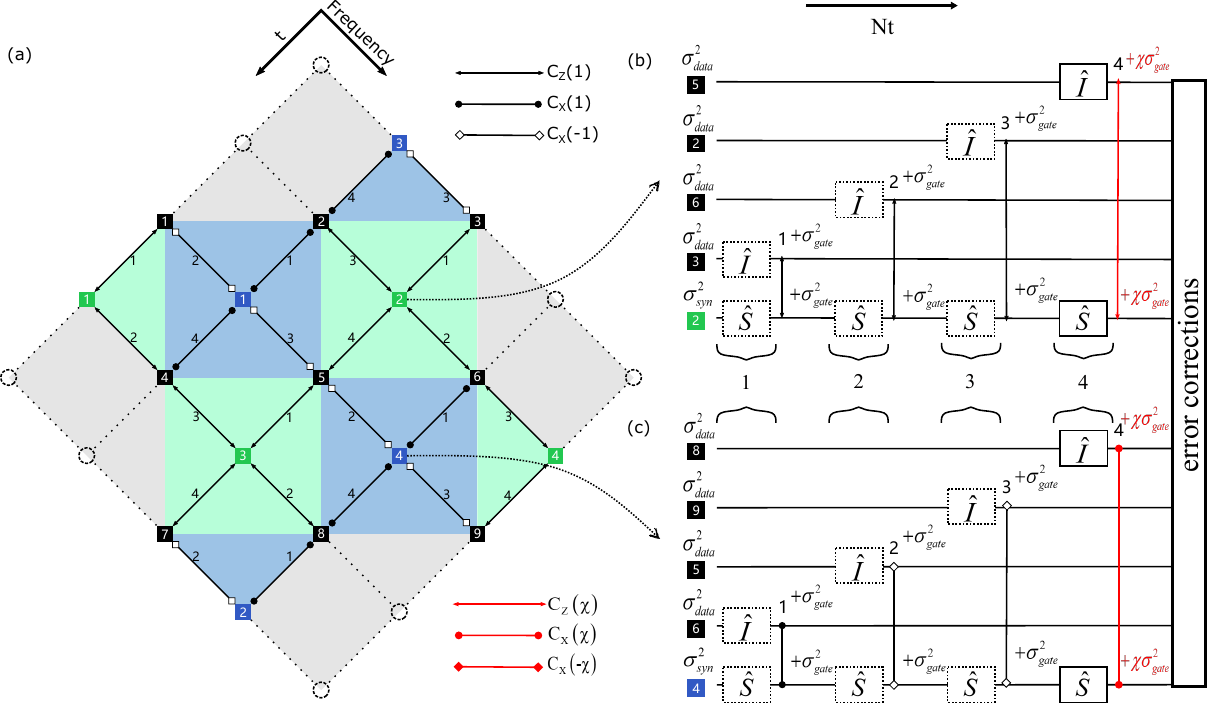}
\caption{The partially squeezed surface-GKP codes. (a) Code layout for distance $d=3$. (b) Single-round Z-type stabilizer measurement. (c) Single-round X-type stabilizer measurement. Black squares denote data GKP qubits, while green and blue squares correspond to Z-type and X-type syndrome GKP qubits, respectively. Dashed circles represent idler modes, which can be erased by measuring the quadrature amplitude without affecting the surface code. Black and red lines represent two-mode gates with weight 1 and $\chi$, respectively.}
\label{Fig5}
\end{figure}
\begin{figure*}[t]
   \centering
    \subfigure[]{\includegraphics[height=0.18\textheight,width=0.32
    \linewidth]{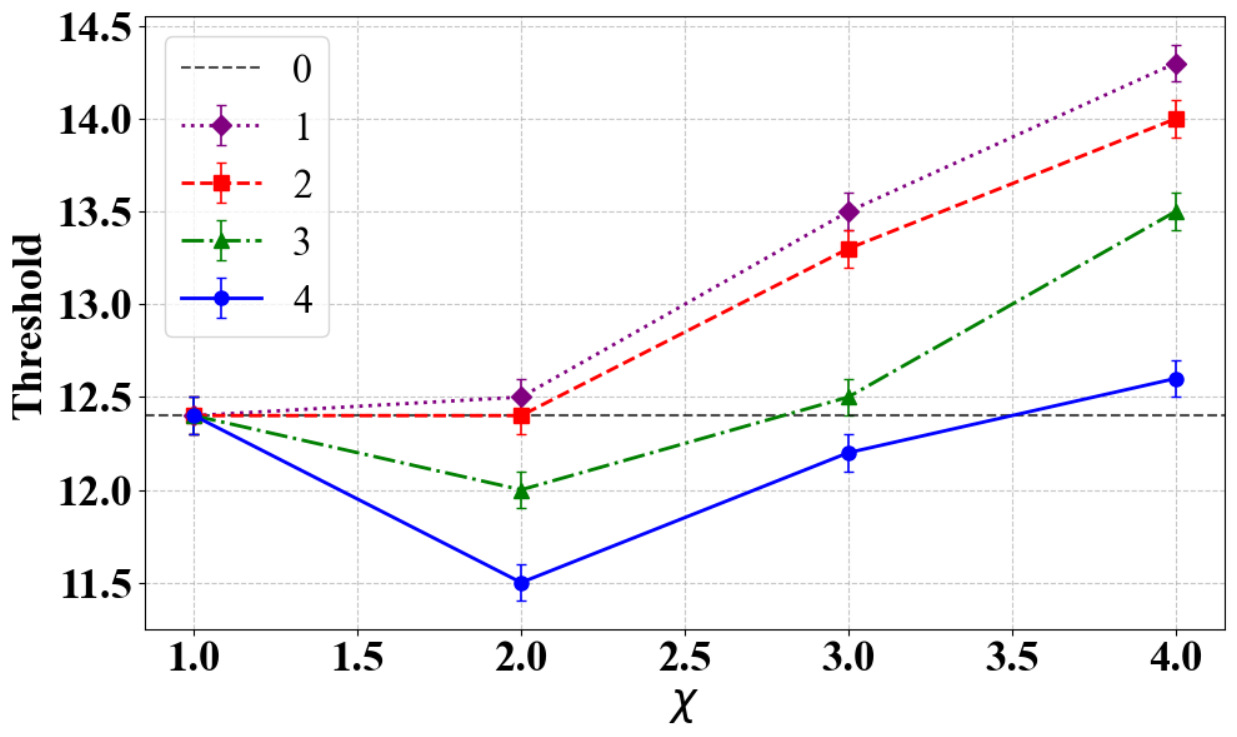}}
    \subfigure[]{\includegraphics[height=0.18\textheight,width=0.32
    \linewidth]{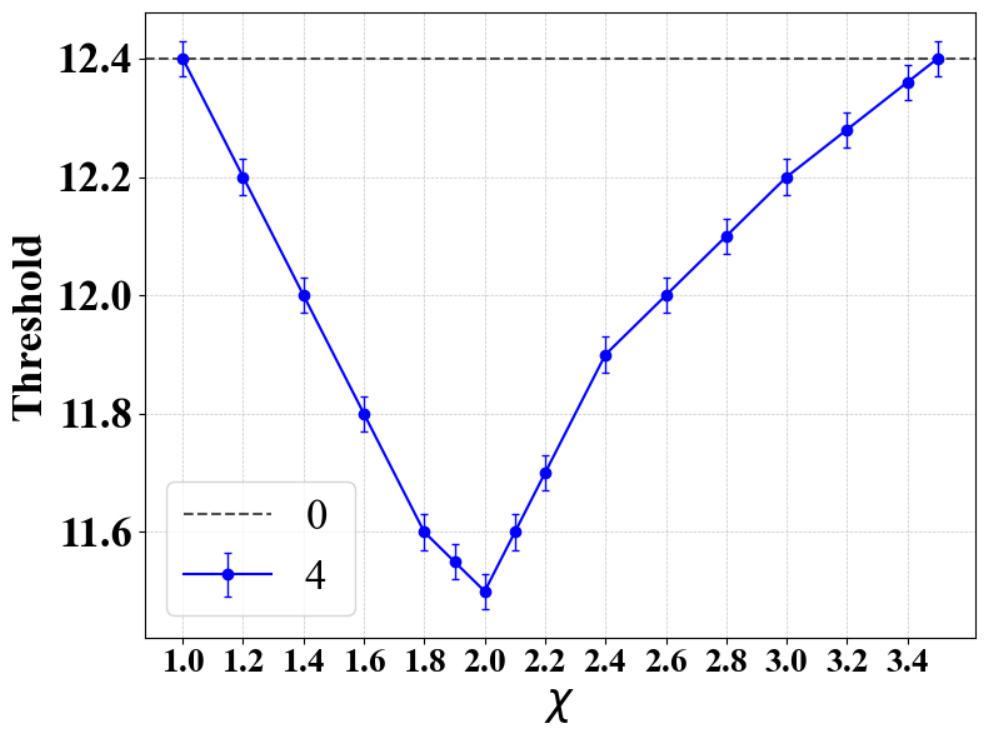}}\subfigure[]{\includegraphics[height=0.18\textheight,width=0.32\linewidth]{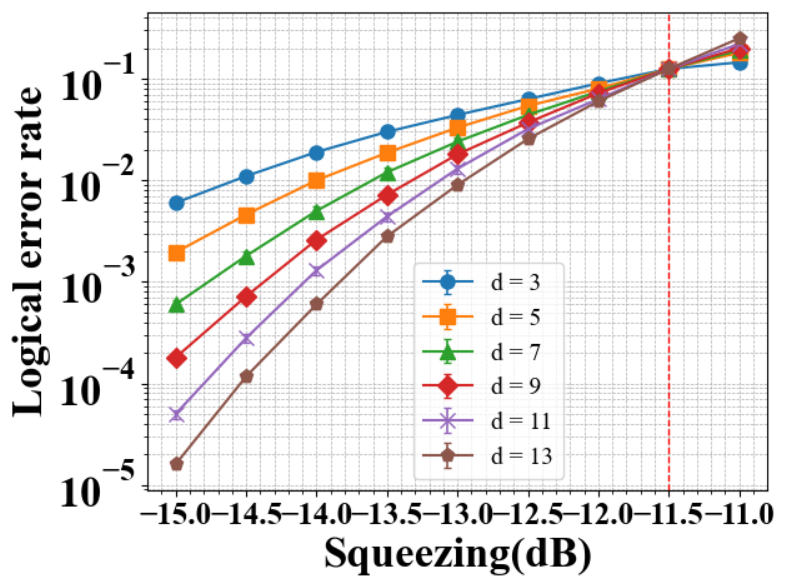}}
    \caption{The fault-tolerant performance of the partially squeezed surface-GKP code. (a) Comparison of introducing squeezing gate implementations in different steps. Curve 0 denotes the standard surface-GKP code, while curves 1–4 correspond to implementating squeezed gates in stabilizer measurement steps 1–4, respectively. (b) Detailed view of curve 4 in (a). (c) Logical error rates for different code distances with $\chi=2$. The red dashed line marks the fault-tolerant squeezing threshold of 11.5 dB.}
    \label{Fig6}
\end{figure*}

\subsection{The partially squeezed surface-GKP code}

Based on our generated 3D cluster state shown in Fig.3(d), we take the surface code with distance $d=3$ as an example, whose data and syndrome GKP qubits are depicted in Fig.\ref{Fig5}(a). There are $d^{2}$ data GKP qubits (black squares) and $\left(d^{2}-1\right)$ syndrome GKP qubits (half Z-type (green squares) and half X-type (blue squares)). Specifically, Z-type and X-type syndrome GKP qubits are prepared in $\left|+\right\rangle_{\mathrm{gkp}}$ and $\left|0\right\rangle_{\mathrm{gkp}}$, respectively, where $\left|+\right\rangle_{\mathrm{gkp}}=\sum_{n\in Z}\left|p=2n\sqrt{\pi}\right\rangle$ and $\left|0\right\rangle_{\mathrm{gkp}}=\sum_{n\in Z}\left|x=2n\sqrt{\pi}\right\rangle$. During Z-type stabilizer measurements (green areas), Z-type syndrome GKP qubits are coupled to data GKP qubits via $\hat{C}_{\mathrm{Z}}\left(1\right)$ to detect Pauli X errors. For X-type stabilizer measurements (blue areas), X-type syndrome GKP qubits are coupled to data GKP qubits via $\hat{C}_{\mathrm{X}}\left(\pm 1\right)$ to detect Pauli Z errors. Each round of Z-type and X-type stabilizer measurement involves four two-mode gate implementation steps (corresponding to the numbers 1, 2, 3, and 4 in Fig.\ref{Fig5}(a)) in $Nt$ direction, as shown in Figs.\ref{Fig5}(b) and \ref{Fig5}(c), respectively. Although the residual Pauli errors can be corrected after final error correction, the overall error rate may also increased due to the accumulation of gate noise during the two-mode gate implementations, which can be modeled by a Gaussian random shift error with variance $\sigma_{\mathrm{gate}}^{2}$\cite{du2025complete}.

To solve this problem, we present the partially squeezed surface-GKP code, introducing a squeezing gate implementation only for syndrome GKP qubits before the two-mode gate implementation in a certain step of stabilizer measurement. It is quite different from other asymmetric surface-GKP codes that require the initial qubits to be biased GKP states\cite{biassurface,XZZXsurfacecode,KyungjooLow-Overhead}. The single-mode squeezing gate $\hat{S}$ is defined as:
\begin{equation}
    \hat{S}\left(\chi\right)=e^{\frac{1}{2}\left(\chi \hat{b}^{2}-\chi^{*}\hat{b}^{\dagger 2}\right)} 
\end{equation}
where $\chi$ is the squeezing parameter. When the single-mode squeezing gate $\hat{S}$ is acting on the syndrome GKP qubit, the single-mode identity gate $\hat{I}$ is applied to the data GKP qubit to maintain computational synchronization. These single-mode gates can be implemented by using the single-mode gate operation shown by the light green area in Fig.3(d). 

The computational basis states of squeezed GKP qubits are defined as
\begin{eqnarray}
\begin{aligned}
&\hat{S}\left(\chi\right)\left|0\right\rangle_{gkp}=\sum_{n\in Z}\left|x=2n\chi\sqrt{\pi}\right\rangle,   \\ 
&\hat{S}\left(\chi\right)\left|1\right\rangle_{gkp}=\sum_{n\in Z}\left|x=\left(2n+1\right)\chi\sqrt{\pi}\right\rangle,   
\end{aligned}
\end{eqnarray}
and the complementary basis states are given by
\begin{eqnarray}
\begin{aligned}
&\hat{S}\left(1/\chi\right)\left|+\right\rangle_{gkp}=\sum_{n\in Z}\left|p=2n\chi\sqrt{\pi}\right\rangle,   \\ 
&\hat{S}\left(1/\chi\right)\left|-\right\rangle_{gkp}=\sum_{n\in Z}\left|p=\left(2n+1\right)\chi\sqrt{\pi}\right\rangle.
\end{aligned}
\end{eqnarray}
The squeezing parameters that are implemented to Z-type and X-type syndrome GKP states are $\chi$ and $1/\chi$, respectively. It will back to the standard surface-GKP case if the squeezing parameter is 1. Taking $\chi>1$, for the Z-type and X-type syndrome GKP qubits, the probability of Pauli X errors and Pauli Z errors can be effectively decreased, respectively. After the squeezing gate implementations, the consequent two-mode $\hat{C}_{\mathrm{Z}}$ and $\hat{C}_{\mathrm{X}}$ gates should be correspondingly scaled as:
\begin{eqnarray}
\begin{aligned}
&\hat{C}_{\mathrm{Z}}\left(\pm \chi\right)=e^{\pm i\chi\hat{x}_{c}\hat{x}_{t}},  \\ 
&\hat{C}_{\mathrm{X}}\left(\pm \chi\right)=e^{\pm i\chi\hat{p}_{c}\hat{p}_{t}},  
\end{aligned}
\end{eqnarray}
where the subscript $c$ and $t$ represent the control and target qubits, respectively. After the scaled $\hat{C}_{\mathrm{X}}\left(\pm \chi\right)$ or $\hat{C}_{\mathrm{Z}}\left(\pm \chi\right)$ gates, the variance $\sigma_{\mathrm{gate}}^{2}$ of the added two-mode gate noise is also scaled to $\chi \sigma_{\mathrm{gate}}^{2}$. Therefore, introducing the partially squeezed surface-GKP code can decrease the error rate in the cost of increasing the gate noise in a certain step of the stabilizer measurement. Next, we will find the optimal trade-off between these two points.

\subsection{The fault-tolerant performance of introducing squeezing gate implementations in different steps}

In principle, one can choose to implement the squeezing gates in any of the four stabilizer measurement steps, i.e., any vertical pair of $\hat{S}$ and $\hat{I}$ operators shown in Fig.5(b) and Fig.5(c). Thus, it is necessary to find which one is the optimal choice by comparing the fault-tolerant performance of introducing squeezed gate implementation in steps 1, 2, 3, and 4. The corresponding numerical simulations are depicted in Fig.6, based on the Monte Carlo method\cite{Gidney2021stim,higgott2022pymatching} where each point is obtained from 100,000 simulations. And the minimum-weight perfect matching decoder is employed to decode the surface code.

For precision, full circuit-level noise model is considered, including the shift errors in the preparation of syndrome GKP qubits (with variance $\sigma_{\mathrm{syn}}^{2}$ in $\hat{x}$ and $\hat{p}$ quadratures), the two-mode gate noise (with variance $\sigma_{\mathrm{gate}}^{2}$ in $\hat{x}$ and $\hat{p}$ quadratures) during the stabilizer measurement, and the residual shift errors in data qubits (with variance $\sigma_{\mathrm{data}}^{2}=\sigma_{\mathrm{gkp}}^{2}+\sigma_{\mathrm{gate}}^{2}$ in $\hat{x}$ and $\hat{p}$ quadratures) after teleportation-based GKP error correction. We assume that $\sigma_{\mathrm{syn}}^{2}=\sigma_{\mathrm{gkp}}^{2}=\sigma_{\mathrm{gate}}^{2}=e^{-2r}/2$, since they come from the same resource preparation. 

How the squeezing parameter value affect the fault-tolerant squeezing threshold is shown in Fig.\ref{Fig6}(a), with implementing squeezed gate in step 1, 2, 3, or 4. For implementing squeezing gates in stabilizer measurement step 1 or 2, the fault-tolerant squeezing threshold is not lower than that of using the standard surface-GKP code (12.4 dB, denoted by "step 0"), and the fault-tolerant performance is even worse for those squeezing parameters bigger than 2. In the later stabilizer measurement steps after implementing the squeezed gate, the two-mode gate becomes $(\hat{C}_{\mathrm{Z}}\left(\chi\right),\hat{C}_{\mathrm{X}}\left(\pm\chi\right))$ and the corresponding variance of gate niose becomes $\chi\sigma_{\mathrm{gate}}^{2}$. Thus, the earlier the implementation of squeezing gates, the more the accumulated gate noise. The increasing gate noise dominates the fault-tolerant process and lead to a increasing overall error rate based on the partially squeezed surface-GKP code. For implementing squeezing gates in stabilizer measurement step 3 and 4, the fault-tolerant squeezing threshold is evidently lower than that of using the standard surface-GKP code, and the fault-tolerant performance is pretty well for squeezing parameter range $\chi\in\left(1,2.7\right)$ and $\chi\in\left(1,3.5\right)$, respectively. The optimal squeezing parameter to achieve the minimum fault-tolerant squeezing threshold is $\chi=2$ for both conditions. Here, the positive effect of implementing the squeezing gates during the stabilizer measurement (reducing the error rate of the syndrome GKP qubits) dominates to give a better fault-tolerant performance, due to less involved gate noise with variance $\chi\sigma_{\mathrm{gate}}^{2}$. It is obvious that implementing squeezing gates in stabilizer measurement step 4 is the best choice, so we use solid-line box to represent the corresponding $\hat{S}$ and $\hat{I}$ operations in step 4 and dotted-line box for that in other steps. A detailed result for the optimized case (implementing squeezing gates in stabilizer measurement step 4) in squeezing parameter range $\chi\in\left(1,3.5\right)$ is shown in Fig.\ref{Fig6}(b). It is clear that the fault-tolerant squeezing threshold reaches 11.5 dB for squeezing parameter $\chi=2$. Based on the partially squeezed surface-GKP code with implementing squeezing gates in stabilizer measurement step 4 and taking squeezing parameter $\chi=2$, the logical error rates as functions of squeezing for different surface code distances $d=3, 5, 7, 9, 11, 13$ are shown in Fig.\ref{Fig6}(c). It is found that the larger the surface code distance, the lower the logical error rate. At the maximum achievable squeezing level of squeezed light (15 dB), the final logical error rate can be reduced to nearly $10^{-5}$ with code distance $d=13$. It can be inferred that the error rate can be further reduced by choosing a larger code distance.

\section{Conclusion}

In this work, we demonstrate a GKP-state-embedded FTQC architecture based on the 3D cluster state generated by our well-designed OEGs. Specifically, the GKP states can be generated deterministically and utilized conveniently for error-correction and fault-tolerance. Moreover, we propose the partially squeezed surface-GKP code by implementing the squeezing gates in the latest stabilizer measurement step. With a squeezing parameter $\chi=2$, the fault-tolerant squeezing threshold can be reduced to 11.5 dB. At the maximum achievable squeezing level (15 dB), the final logical error rate can be lowered to nearly $10^{-5}$ when the code distance is $d=13$, and this error rate can be further reduced by increasing the code distance. Additionally, the fault-tolerant squeezing threshold and logical error rate can be further improved by considering analog information from the GKP error correction\cite{AnalogQuantumErrorPhysRevLett} or constructing 3D cluster states compatible with quantum low-density parity-check codes\cite{ArbitraryPRL}. Our scheme is flexible: diverse 3D entanglement structures can be achieved by adopting different combinations of BSs and time delays, and GKP states can be injected into the cluster state for on-demand utilization. In a word, our work offers valuable insights for the experimental realization of FTQC. 

\section*{acknowledgments}
This work was supported by the Innovation Program for Quantum Science and Technology (Grant No. 2023ZD0300400), the National Key Research and Development Program of China (Grant No. 2021YFC2201802), and the Central Government Guidance Funds for Local Science and Technology Development Projects (Grant No. YDZJSX2025D001).

\nocite{*}
\bibliography{main}% Produces the bibliography via BibTeX.

\begin{thebibliography}{41}%
\makeatletter
\providecommand \@ifxundefined [1]{%
 \@ifx{#1\undefined}
}%
\providecommand \@ifnum [1]{%
 \ifnum #1\expandafter \@firstoftwo
 \else \expandafter \@secondoftwo
 \fi
}%
\providecommand \@ifx [1]{%
 \ifx #1\expandafter \@firstoftwo
 \else \expandafter \@secondoftwo
 \fi
}%
\providecommand \natexlab [1]{#1}%
\providecommand \enquote  [1]{``#1''}%
\providecommand \bibnamefont  [1]{#1}%
\providecommand \bibfnamefont [1]{#1}%
\providecommand \citenamefont [1]{#1}%
\providecommand \href@noop [0]{\@secondoftwo}%
\providecommand \href [0]{\begingroup \@sanitize@url \@href}%
\providecommand \@href[1]{\@@startlink{#1}\@@href}%
\providecommand \@@href[1]{\endgroup#1\@@endlink}%
\providecommand \@sanitize@url [0]{\catcode `\\12\catcode `\$12\catcode `\&12\catcode `\#12\catcode `\^12\catcode `\_12\catcode `\%12\relax}%
\providecommand \@@startlink[1]{}%
\providecommand \@@endlink[0]{}%
\providecommand \url  [0]{\begingroup\@sanitize@url \@url }%
\providecommand \@url [1]{\endgroup\@href {#1}{\urlprefix }}%
\providecommand \urlprefix  [0]{URL }%
\providecommand \Eprint [0]{\href }%
\providecommand \doibase [0]{https://doi.org/}%
\providecommand \selectlanguage [0]{\@gobble}%
\providecommand \bibinfo  [0]{\@secondoftwo}%
\providecommand \bibfield  [0]{\@secondoftwo}%
\providecommand \translation [1]{[#1]}%
\providecommand \BibitemOpen [0]{}%
\providecommand \bibitemStop [0]{}%
\providecommand \bibitemNoStop [0]{.\EOS\space}%
\providecommand \EOS [0]{\spacefactor3000\relax}%
\providecommand \BibitemShut  [1]{\csname bibitem#1\endcsname}%
\let\auto@bib@innerbib\@empty
%</preamble>
\bibitem [{\citenamefont {Rad}\ \emph {et~al.}(2025)\citenamefont {Rad}, \citenamefont {Ainsworth}, \citenamefont {Alexander}, \citenamefont {Altieri}, \citenamefont {Askarani}, \citenamefont {Baby}, \citenamefont {Banchi}, \citenamefont {Baragiola}, \citenamefont {Bourassa},\ and\ \citenamefont {Chadwick}}]{2025Scaling}%
  \BibitemOpen
  \bibfield  {author} {\bibinfo {author} {\bibfnamefont {H.~A.}\ \bibnamefont {Rad}}, \bibinfo {author} {\bibfnamefont {T.}~\bibnamefont {Ainsworth}}, \bibinfo {author} {\bibfnamefont {R.~N.}\ \bibnamefont {Alexander}}, \bibinfo {author} {\bibfnamefont {B.}~\bibnamefont {Altieri}}, \bibinfo {author} {\bibfnamefont {M.~F.}\ \bibnamefont {Askarani}}, \bibinfo {author} {\bibfnamefont {R.}~\bibnamefont {Baby}}, \bibinfo {author} {\bibfnamefont {L.}~\bibnamefont {Banchi}}, \bibinfo {author} {\bibfnamefont {B.~Q.}\ \bibnamefont {Baragiola}}, \bibinfo {author} {\bibfnamefont {J.~E.}\ \bibnamefont {Bourassa}},\ and\ \bibinfo {author} {\bibfnamefont {R.~S.}\ \bibnamefont {Chadwick}},\ }\bibfield  {title} {\bibinfo {title} {Scaling and networking a modular photonic quantum computer},\ }\href@noop {} {\bibfield  {journal} {\bibinfo  {journal} {Nature}\ }\textbf {\bibinfo {volume} {638}},\ \bibinfo {pages} {912} (\bibinfo {year} {2025})}\BibitemShut {NoStop}%
\bibitem [{\citenamefont {Alexander}\ \emph {et~al.}(2017)\citenamefont {Alexander}, \citenamefont {Gabay}, \citenamefont {Rohde},\ and\ \citenamefont {Menicucci}}]{universalityPRL}%
  \BibitemOpen
  \bibfield  {author} {\bibinfo {author} {\bibfnamefont {R.~N.}\ \bibnamefont {Alexander}}, \bibinfo {author} {\bibfnamefont {N.~C.}\ \bibnamefont {Gabay}}, \bibinfo {author} {\bibfnamefont {P.~P.}\ \bibnamefont {Rohde}},\ and\ \bibinfo {author} {\bibfnamefont {N.~C.}\ \bibnamefont {Menicucci}},\ }\bibfield  {title} {\bibinfo {title} {Measurement-based linear optics},\ }\href {https://doi.org/10.1103/PhysRevLett.118.110503} {\bibfield  {journal} {\bibinfo  {journal} {Phys. Rev. Lett.}\ }\textbf {\bibinfo {volume} {118}},\ \bibinfo {pages} {110503} (\bibinfo {year} {2017})}\BibitemShut {NoStop}%
\bibitem [{\citenamefont {Asavanant}\ and\ \citenamefont {Furusawa}(2024)}]{Furusawa2024review}%
  \BibitemOpen
  \bibfield  {author} {\bibinfo {author} {\bibfnamefont {W.}~\bibnamefont {Asavanant}}\ and\ \bibinfo {author} {\bibfnamefont {A.}~\bibnamefont {Furusawa}},\ }\bibfield  {title} {\bibinfo {title} {Multipartite continuous-variable optical quantum entanglement: Generation and application},\ }\href {https://doi.org/10.1103/PhysRevA.109.040101} {\bibfield  {journal} {\bibinfo  {journal} {Phys. Rev. A}\ }\textbf {\bibinfo {volume} {109}},\ \bibinfo {pages} {040101} (\bibinfo {year} {2024})}\BibitemShut {NoStop}%
\bibitem [{\citenamefont {Fukui}\ \emph {et~al.}(2020)\citenamefont {Fukui}, \citenamefont {Asavanant},\ and\ \citenamefont {Furusawa}}]{Fukui3D2020}%
  \BibitemOpen
  \bibfield  {author} {\bibinfo {author} {\bibfnamefont {K.}~\bibnamefont {Fukui}}, \bibinfo {author} {\bibfnamefont {W.}~\bibnamefont {Asavanant}},\ and\ \bibinfo {author} {\bibfnamefont {A.}~\bibnamefont {Furusawa}},\ }\bibfield  {title} {\bibinfo {title} {Temporal-mode continuous-variable three-dimensional cluster state for topologically protected measurement-based quantum computation},\ }\href {https://doi.org/10.1103/PhysRevA.102.032614} {\bibfield  {journal} {\bibinfo  {journal} {Phys. Rev. A}\ }\textbf {\bibinfo {volume} {102}},\ \bibinfo {pages} {032614} (\bibinfo {year} {2020})}\BibitemShut {NoStop}%
\bibitem [{\citenamefont {Zhu}\ \emph {et~al.}(2021)\citenamefont {Zhu}, \citenamefont {Chang}, \citenamefont {Gonz\'{a}lez-Arciniegas}, \citenamefont {Pe'er}, \citenamefont {Higgins},\ and\ \citenamefont {Pfister}}]{Oliver3D2021}%
  \BibitemOpen
  \bibfield  {author} {\bibinfo {author} {\bibfnamefont {X.}~\bibnamefont {Zhu}}, \bibinfo {author} {\bibfnamefont {C.}~\bibnamefont {Chang}}, \bibinfo {author} {\bibfnamefont {C.}~\bibnamefont {Gonz\'{a}lez-Arciniegas}}, \bibinfo {author} {\bibfnamefont {A.}~\bibnamefont {Pe'er}}, \bibinfo {author} {\bibfnamefont {J.}~\bibnamefont {Higgins}},\ and\ \bibinfo {author} {\bibfnamefont {O.}~\bibnamefont {Pfister}},\ }\bibfield  {title} {\bibinfo {title} {Hypercubic cluster states in the phase-modulated quantum optical frequency comb},\ }\href {https://doi.org/10.1364/OPTICA.411713} {\bibfield  {journal} {\bibinfo  {journal} {Optica}\ }\textbf {\bibinfo {volume} {8}},\ \bibinfo {pages} {281} (\bibinfo {year} {2021})}\BibitemShut {NoStop}%
\bibitem [{\citenamefont {Wu}\ \emph {et~al.}(2020)\citenamefont {Wu}, \citenamefont {Alexander}, \citenamefont {Liu},\ and\ \citenamefont {Zhang}}]{wu(2020)}%
  \BibitemOpen
  \bibfield  {author} {\bibinfo {author} {\bibfnamefont {B.}~\bibnamefont {Wu}}, \bibinfo {author} {\bibfnamefont {R.~N.}\ \bibnamefont {Alexander}}, \bibinfo {author} {\bibfnamefont {S.}~\bibnamefont {Liu}},\ and\ \bibinfo {author} {\bibfnamefont {Z.}~\bibnamefont {Zhang}},\ }\bibfield  {title} {\bibinfo {title} {Quantum computing with multidimensional continuous-variable cluster states in a scalable photonic platform},\ }\href@noop {} {\bibfield  {journal} {\bibinfo  {journal} {Phys. Rev. Research}\ }\textbf {\bibinfo {volume} {2}},\ \bibinfo {pages} {023138} (\bibinfo {year} {2020})}\BibitemShut {NoStop}%
\bibitem [{\citenamefont {Du}\ \emph {et~al.}(2023)\citenamefont {Du}, \citenamefont {Wang}, \citenamefont {Liu}, \citenamefont {Yang},\ and\ \citenamefont {Zhang}}]{Du(2022)}%
  \BibitemOpen
  \bibfield  {author} {\bibinfo {author} {\bibfnamefont {P.}~\bibnamefont {Du}}, \bibinfo {author} {\bibfnamefont {Y.}~\bibnamefont {Wang}}, \bibinfo {author} {\bibfnamefont {K.}~\bibnamefont {Liu}}, \bibinfo {author} {\bibfnamefont {R.}~\bibnamefont {Yang}},\ and\ \bibinfo {author} {\bibfnamefont {J.}~\bibnamefont {Zhang}},\ }\bibfield  {title} {\bibinfo {title} {Generation of large-scale continuous-variable cluster states multiplexed both in time and frequency domains},\ }\href {https://doi.org/10.1364/OE.479420} {\bibfield  {journal} {\bibinfo  {journal} {Opt. Express}\ }\textbf {\bibinfo {volume} {31}},\ \bibinfo {pages} {7535} (\bibinfo {year} {2023})}\BibitemShut {NoStop}%
\bibitem [{\citenamefont {Larsen}\ \emph {et~al.}(2020)\citenamefont {Larsen}, \citenamefont {Neergaard-Nielsen},\ and\ \citenamefont {Andersen}}]{Larsen(2020)noiseanalysis}%
  \BibitemOpen
  \bibfield  {author} {\bibinfo {author} {\bibfnamefont {M.~V.}\ \bibnamefont {Larsen}}, \bibinfo {author} {\bibfnamefont {J.~S.}\ \bibnamefont {Neergaard-Nielsen}},\ and\ \bibinfo {author} {\bibfnamefont {U.~L.}\ \bibnamefont {Andersen}},\ }\bibfield  {title} {\bibinfo {title} {Architecture and noise analysis of continuous-variable quantum gates using two-dimensional cluster states},\ }\href {https://doi.org/10.1103/PhysRevA.102.042608} {\bibfield  {journal} {\bibinfo  {journal} {Phys. Rev. A}\ }\textbf {\bibinfo {volume} {102}},\ \bibinfo {pages} {042608} (\bibinfo {year} {2020})}\BibitemShut {NoStop}%
\bibitem [{\citenamefont {Alexander}\ \emph {et~al.}(2014)\citenamefont {Alexander}, \citenamefont {Armstrong}, \citenamefont {Ukai},\ and\ \citenamefont {Menicucci}}]{alexander2014noise}%
  \BibitemOpen
  \bibfield  {author} {\bibinfo {author} {\bibfnamefont {R.~N.}\ \bibnamefont {Alexander}}, \bibinfo {author} {\bibfnamefont {S.~C.}\ \bibnamefont {Armstrong}}, \bibinfo {author} {\bibfnamefont {R.}~\bibnamefont {Ukai}},\ and\ \bibinfo {author} {\bibfnamefont {N.~C.}\ \bibnamefont {Menicucci}},\ }\bibfield  {title} {\bibinfo {title} {Noise analysis of single-mode gaussian operations using continuous-variable cluster states},\ }\href@noop {} {\bibfield  {journal} {\bibinfo  {journal} {Phys. Rev. A}\ }\textbf {\bibinfo {volume} {90}},\ \bibinfo {pages} {062324} (\bibinfo {year} {2014})}\BibitemShut {NoStop}%
\bibitem [{\citenamefont {Du}\ \emph {et~al.}(2025)\citenamefont {Du}, \citenamefont {Zhang}, \citenamefont {Zhang}, \citenamefont {Yang},\ and\ \citenamefont {Gao}}]{du2025complete}%
  \BibitemOpen
  \bibfield  {author} {\bibinfo {author} {\bibfnamefont {P.}~\bibnamefont {Du}}, \bibinfo {author} {\bibfnamefont {J.}~\bibnamefont {Zhang}}, \bibinfo {author} {\bibfnamefont {T.}~\bibnamefont {Zhang}}, \bibinfo {author} {\bibfnamefont {R.}~\bibnamefont {Yang}},\ and\ \bibinfo {author} {\bibfnamefont {J.}~\bibnamefont {Gao}},\ }\bibfield  {title} {\bibinfo {title} {A complete continuous-variable quantum computation architecture based on the 2d spatiotemporal cluster state},\ }\href@noop {} {\bibfield  {journal} {\bibinfo  {journal} {Scientific Reports}\ }\textbf {\bibinfo {volume} {15}},\ \bibinfo {pages} {18199} (\bibinfo {year} {2025})}\BibitemShut {NoStop}%
\bibitem [{\citenamefont {Gottesman}\ \emph {et~al.}(2000)\citenamefont {Gottesman}, \citenamefont {Kitaev},\ and\ \citenamefont {Preskill}}]{Gottesman(2000)GKPcoding}%
  \BibitemOpen
  \bibfield  {author} {\bibinfo {author} {\bibfnamefont {D.}~\bibnamefont {Gottesman}}, \bibinfo {author} {\bibfnamefont {A.~Y.}\ \bibnamefont {Kitaev}},\ and\ \bibinfo {author} {\bibfnamefont {J.}~\bibnamefont {Preskill}},\ }\bibfield  {title} {\bibinfo {title} {Encoding a qubit in an oscillator},\ }\href@noop {} {\bibfield  {journal} {\bibinfo  {journal} {Phys. Rev. A}\ }\textbf {\bibinfo {volume} {64}},\ \bibinfo {pages} {012310} (\bibinfo {year} {2000})}\BibitemShut {NoStop}%
\bibitem [{\citenamefont {Noh}\ \emph {et~al.}(2020)\citenamefont {Noh}, \citenamefont {Girvin},\ and\ \citenamefont {Jiang}}]{GKPPRL}%
  \BibitemOpen
  \bibfield  {author} {\bibinfo {author} {\bibfnamefont {K.}~\bibnamefont {Noh}}, \bibinfo {author} {\bibfnamefont {S.~M.}\ \bibnamefont {Girvin}},\ and\ \bibinfo {author} {\bibfnamefont {L.}~\bibnamefont {Jiang}},\ }\bibfield  {title} {\bibinfo {title} {Encoding an oscillator into many oscillators},\ }\href {https://doi.org/10.1103/PhysRevLett.125.080503} {\bibfield  {journal} {\bibinfo  {journal} {Phys. Rev. Lett.}\ }\textbf {\bibinfo {volume} {125}},\ \bibinfo {pages} {080503} (\bibinfo {year} {2020})}\BibitemShut {NoStop}%
\bibitem [{\citenamefont {Bourassa}\ \emph {et~al.}(2021)\citenamefont {Bourassa}, \citenamefont {Alexander}, \citenamefont {Vasmer}, \citenamefont {Patil}, \citenamefont {Tzitrin}, \citenamefont {Matsuura}, \citenamefont {Su}, \citenamefont {Baragiola}, \citenamefont {Guha}, \citenamefont {Dauphinais}, \citenamefont {Sabapathy}, \citenamefont {Menicucci},\ and\ \citenamefont {Dhand}}]{Bour2021blue}%
  \BibitemOpen
  \bibfield  {author} {\bibinfo {author} {\bibfnamefont {J.~E.}\ \bibnamefont {Bourassa}}, \bibinfo {author} {\bibfnamefont {R.~N.}\ \bibnamefont {Alexander}}, \bibinfo {author} {\bibfnamefont {M.}~\bibnamefont {Vasmer}}, \bibinfo {author} {\bibfnamefont {A.}~\bibnamefont {Patil}}, \bibinfo {author} {\bibfnamefont {I.}~\bibnamefont {Tzitrin}}, \bibinfo {author} {\bibfnamefont {T.}~\bibnamefont {Matsuura}}, \bibinfo {author} {\bibfnamefont {D.}~\bibnamefont {Su}}, \bibinfo {author} {\bibfnamefont {B.~Q.}\ \bibnamefont {Baragiola}}, \bibinfo {author} {\bibfnamefont {S.}~\bibnamefont {Guha}}, \bibinfo {author} {\bibfnamefont {G.}~\bibnamefont {Dauphinais}}, \bibinfo {author} {\bibfnamefont {K.~K.}\ \bibnamefont {Sabapathy}}, \bibinfo {author} {\bibfnamefont {N.~C.}\ \bibnamefont {Menicucci}},\ and\ \bibinfo {author} {\bibfnamefont {I.}~\bibnamefont {Dhand}},\ }\bibfield  {title} {\bibinfo {title} {Blueprint for a {S}calable {P}hotonic {F}ault-{T}olerant {Q}uantum {C}omputer},\ }\href
  {https://doi.org/10.22331/q-2021-02-04-392} {\bibfield  {journal} {\bibinfo  {journal} {{Quantum}}\ }\textbf {\bibinfo {volume} {5}},\ \bibinfo {pages} {392} (\bibinfo {year} {2021})}\BibitemShut {NoStop}%
\bibitem [{\citenamefont {Noh}\ \emph {et~al.}(2022)\citenamefont {Noh}, \citenamefont {Chamberland},\ and\ \citenamefont {Brand\~ao}}]{KyungjooLow-Overhead}%
  \BibitemOpen
  \bibfield  {author} {\bibinfo {author} {\bibfnamefont {K.}~\bibnamefont {Noh}}, \bibinfo {author} {\bibfnamefont {C.}~\bibnamefont {Chamberland}},\ and\ \bibinfo {author} {\bibfnamefont {F.~G.}\ \bibnamefont {Brand\~ao}},\ }\bibfield  {title} {\bibinfo {title} {Low-overhead fault-tolerant quantum error correction with the surface-gkp code},\ }\href@noop {} {\bibfield  {journal} {\bibinfo  {journal} {PRX Quantum}\ }\textbf {\bibinfo {volume} {3}},\ \bibinfo {pages} {010315} (\bibinfo {year} {2022})}\BibitemShut {NoStop}%
\bibitem [{\citenamefont {Dahan}\ \emph {et~al.}(2023)\citenamefont {Dahan}, \citenamefont {Baranes}, \citenamefont {Gorlach}, \citenamefont {Ruimy}, \citenamefont {Rivera},\ and\ \citenamefont {Kaminer}}]{DahanGENGKP}%
  \BibitemOpen
  \bibfield  {author} {\bibinfo {author} {\bibfnamefont {R.}~\bibnamefont {Dahan}}, \bibinfo {author} {\bibfnamefont {G.}~\bibnamefont {Baranes}}, \bibinfo {author} {\bibfnamefont {A.}~\bibnamefont {Gorlach}}, \bibinfo {author} {\bibfnamefont {R.}~\bibnamefont {Ruimy}}, \bibinfo {author} {\bibfnamefont {N.}~\bibnamefont {Rivera}},\ and\ \bibinfo {author} {\bibfnamefont {I.}~\bibnamefont {Kaminer}},\ }\bibfield  {title} {\bibinfo {title} {Creation of optical cat and gkp states using shaped free electrons},\ }\href@noop {} {\bibfield  {journal} {\bibinfo  {journal} {Phys. Rev. X}\ }\textbf {\bibinfo {volume} {13}},\ \bibinfo {pages} {031001} (\bibinfo {year} {2023})}\BibitemShut {NoStop}%
\bibitem [{\citenamefont {Hastrup}\ and\ \citenamefont {Andersen}(2022)}]{AndersenGENGKP}%
  \BibitemOpen
  \bibfield  {author} {\bibinfo {author} {\bibfnamefont {J.}~\bibnamefont {Hastrup}}\ and\ \bibinfo {author} {\bibfnamefont {U.~L.}\ \bibnamefont {Andersen}},\ }\bibfield  {title} {\bibinfo {title} {Protocol for generating optical gottesman-kitaev-preskill states with cavity qed},\ }\href@noop {} {\bibfield  {journal} {\bibinfo  {journal} {Phys. Rev. Lett.}\ }\textbf {\bibinfo {volume} {128}},\ \bibinfo {pages} {170503} (\bibinfo {year} {2022})}\BibitemShut {NoStop}%
\bibitem [{\citenamefont {Hastrup}\ \emph {et~al.}(2021)\citenamefont {Hastrup}, \citenamefont {Park}, \citenamefont {Brask}, \citenamefont {Filip},\ and\ \citenamefont {Andersen}}]{UlrikGENGKP}%
  \BibitemOpen
  \bibfield  {author} {\bibinfo {author} {\bibfnamefont {J.}~\bibnamefont {Hastrup}}, \bibinfo {author} {\bibfnamefont {K.}~\bibnamefont {Park}}, \bibinfo {author} {\bibfnamefont {J.~B.}\ \bibnamefont {Brask}}, \bibinfo {author} {\bibfnamefont {R.}~\bibnamefont {Filip}},\ and\ \bibinfo {author} {\bibfnamefont {U.~L.}\ \bibnamefont {Andersen}},\ }\bibfield  {title} {\bibinfo {title} {Measurement-free preparation of grid states},\ }\href@noop {} {\bibfield  {journal} {\bibinfo  {journal} {npj Quantum Information}\ }\textbf {\bibinfo {volume} {7}},\ \bibinfo {pages} {17} (\bibinfo {year} {2021})}\BibitemShut {NoStop}%
\bibitem [{\citenamefont {Konno}\ \emph {et~al.}(2024)\citenamefont {Konno}, \citenamefont {Asavanant}, \citenamefont {Hanamura}, \citenamefont {Nagayoshi}, \citenamefont {Fukui}, \citenamefont {Sakaguchi}, \citenamefont {Ide}, \citenamefont {China}, \citenamefont {Yabuno}, \citenamefont {Miki}, \citenamefont {Terai}, \citenamefont {Takase}, \citenamefont {Endo}, \citenamefont {Marek}, \citenamefont {Filip}, \citenamefont {van Loock},\ and\ \citenamefont {Furusawa}}]{FurusawaGKP2025}%
  \BibitemOpen
  \bibfield  {author} {\bibinfo {author} {\bibfnamefont {S.}~\bibnamefont {Konno}}, \bibinfo {author} {\bibfnamefont {W.}~\bibnamefont {Asavanant}}, \bibinfo {author} {\bibfnamefont {F.}~\bibnamefont {Hanamura}}, \bibinfo {author} {\bibfnamefont {H.}~\bibnamefont {Nagayoshi}}, \bibinfo {author} {\bibfnamefont {K.}~\bibnamefont {Fukui}}, \bibinfo {author} {\bibfnamefont {A.}~\bibnamefont {Sakaguchi}}, \bibinfo {author} {\bibfnamefont {R.}~\bibnamefont {Ide}}, \bibinfo {author} {\bibfnamefont {F.}~\bibnamefont {China}}, \bibinfo {author} {\bibfnamefont {M.}~\bibnamefont {Yabuno}}, \bibinfo {author} {\bibfnamefont {S.}~\bibnamefont {Miki}}, \bibinfo {author} {\bibfnamefont {H.}~\bibnamefont {Terai}}, \bibinfo {author} {\bibfnamefont {K.}~\bibnamefont {Takase}}, \bibinfo {author} {\bibfnamefont {M.}~\bibnamefont {Endo}}, \bibinfo {author} {\bibfnamefont {P.}~\bibnamefont {Marek}}, \bibinfo {author} {\bibfnamefont {R.}~\bibnamefont {Filip}}, \bibinfo {author} {\bibfnamefont {P.}~\bibnamefont {van Loock}},\ and\
  \bibinfo {author} {\bibfnamefont {A.}~\bibnamefont {Furusawa}},\ }\bibfield  {title} {\bibinfo {title} {Logical states for fault-tolerant quantum computation with propagating light},\ }\href {https://doi.org/10.1126/science.adk7560} {\bibfield  {journal} {\bibinfo  {journal} {Science}\ }\textbf {\bibinfo {volume} {383}},\ \bibinfo {pages} {289} (\bibinfo {year} {2024})}\BibitemShut {NoStop}%
\bibitem [{\citenamefont {Larsen}\ \emph {et~al.}(2025)\citenamefont {Larsen}, \citenamefont {Bourassa}, \citenamefont {Kocsis}, \citenamefont {Tasker}, \citenamefont {Chadwick}, \citenamefont {Gonz{\'a}lez-Arciniegas}, \citenamefont {Hastrup}, \citenamefont {Lopetegui-Gonz{\'a}lez}, \citenamefont {Miatto}, \citenamefont {Motamedi} \emph {et~al.}}]{larsenGKP2025}%
  \BibitemOpen
  \bibfield  {author} {\bibinfo {author} {\bibfnamefont {M.}~\bibnamefont {Larsen}}, \bibinfo {author} {\bibfnamefont {J.}~\bibnamefont {Bourassa}}, \bibinfo {author} {\bibfnamefont {S.}~\bibnamefont {Kocsis}}, \bibinfo {author} {\bibfnamefont {J.}~\bibnamefont {Tasker}}, \bibinfo {author} {\bibfnamefont {R.}~\bibnamefont {Chadwick}}, \bibinfo {author} {\bibfnamefont {C.}~\bibnamefont {Gonz{\'a}lez-Arciniegas}}, \bibinfo {author} {\bibfnamefont {J.}~\bibnamefont {Hastrup}}, \bibinfo {author} {\bibfnamefont {C.}~\bibnamefont {Lopetegui-Gonz{\'a}lez}}, \bibinfo {author} {\bibfnamefont {F.}~\bibnamefont {Miatto}}, \bibinfo {author} {\bibfnamefont {A.}~\bibnamefont {Motamedi}}, \emph {et~al.},\ }\bibfield  {title} {\bibinfo {title} {Integrated photonic source of gottesman--kitaev--preskill qubits},\ }\href@noop {} {\bibfield  {journal} {\bibinfo  {journal} {Nature}\ ,\ \bibinfo {pages} {1}} (\bibinfo {year} {2025})}\BibitemShut {NoStop}%
\bibitem [{\citenamefont {Larsen}\ \emph {et~al.}(2021)\citenamefont {Larsen}, \citenamefont {Chamberland}, \citenamefont {Noh}, \citenamefont {Neergaard-Nielsen},\ and\ \citenamefont {Andersen}}]{Larsen(2021)PRXQuantum}%
  \BibitemOpen
  \bibfield  {author} {\bibinfo {author} {\bibfnamefont {M.~V.}\ \bibnamefont {Larsen}}, \bibinfo {author} {\bibfnamefont {C.}~\bibnamefont {Chamberland}}, \bibinfo {author} {\bibfnamefont {K.}~\bibnamefont {Noh}}, \bibinfo {author} {\bibfnamefont {J.~S.}\ \bibnamefont {Neergaard-Nielsen}},\ and\ \bibinfo {author} {\bibfnamefont {U.~L.}\ \bibnamefont {Andersen}},\ }\bibfield  {title} {\bibinfo {title} {Fault-tolerant continuous-variable measurement-based quantum computation architecture},\ }\href {https://doi.org/10.1103/PRXQuantum.2.030325} {\bibfield  {journal} {\bibinfo  {journal} {PRX Quantum}\ }\textbf {\bibinfo {volume} {2}},\ \bibinfo {pages} {030325} (\bibinfo {year} {2021})}\BibitemShut {NoStop}%
\bibitem [{\citenamefont {Walshe}\ \emph {et~al.}(2025)\citenamefont {Walshe}, \citenamefont {Baragiola}, \citenamefont {Ferretti}, \citenamefont {Gefaell}, \citenamefont {Vasmer}, \citenamefont {Weil}, \citenamefont {Matsuura}, \citenamefont {Jaeken}, \citenamefont {Pantaleoni}, \citenamefont {Han}, \citenamefont {Hillmann}, \citenamefont {Menicucci}, \citenamefont {Tzitrin},\ and\ \citenamefont {Alexander}}]{ArbitraryPRL}%
  \BibitemOpen
  \bibfield  {author} {\bibinfo {author} {\bibfnamefont {B.~W.}\ \bibnamefont {Walshe}}, \bibinfo {author} {\bibfnamefont {B.~Q.}\ \bibnamefont {Baragiola}}, \bibinfo {author} {\bibfnamefont {H.}~\bibnamefont {Ferretti}}, \bibinfo {author} {\bibfnamefont {J.}~\bibnamefont {Gefaell}}, \bibinfo {author} {\bibfnamefont {M.}~\bibnamefont {Vasmer}}, \bibinfo {author} {\bibfnamefont {R.}~\bibnamefont {Weil}}, \bibinfo {author} {\bibfnamefont {T.}~\bibnamefont {Matsuura}}, \bibinfo {author} {\bibfnamefont {T.}~\bibnamefont {Jaeken}}, \bibinfo {author} {\bibfnamefont {G.}~\bibnamefont {Pantaleoni}}, \bibinfo {author} {\bibfnamefont {Z.}~\bibnamefont {Han}}, \bibinfo {author} {\bibfnamefont {T.}~\bibnamefont {Hillmann}}, \bibinfo {author} {\bibfnamefont {N.~C.}\ \bibnamefont {Menicucci}}, \bibinfo {author} {\bibfnamefont {I.}~\bibnamefont {Tzitrin}},\ and\ \bibinfo {author} {\bibfnamefont {R.~N.}\ \bibnamefont {Alexander}},\ }\bibfield  {title} {\bibinfo {title} {Linear-optical quantum computation with arbitrary
  error-correcting codes},\ }\href {https://doi.org/10.1103/PhysRevLett.134.100602} {\bibfield  {journal} {\bibinfo  {journal} {Phys. Rev. Lett.}\ }\textbf {\bibinfo {volume} {134}},\ \bibinfo {pages} {100602} (\bibinfo {year} {2025})}\BibitemShut {NoStop}%
\bibitem [{\citenamefont {Asavanant}\ \emph {et~al.}(2023)\citenamefont {Asavanant}, \citenamefont {Fukui}, \citenamefont {Sakaguchi},\ and\ \citenamefont {Furusawa}}]{Switching-free}%
  \BibitemOpen
  \bibfield  {author} {\bibinfo {author} {\bibfnamefont {W.}~\bibnamefont {Asavanant}}, \bibinfo {author} {\bibfnamefont {K.}~\bibnamefont {Fukui}}, \bibinfo {author} {\bibfnamefont {A.}~\bibnamefont {Sakaguchi}},\ and\ \bibinfo {author} {\bibfnamefont {A.}~\bibnamefont {Furusawa}},\ }\bibfield  {title} {\bibinfo {title} {Switching-free time-domain optical quantum computation with quantum teleportation},\ }\href@noop {} {\bibfield  {journal} {\bibinfo  {journal} {Phys. Rev. A}\ }\textbf {\bibinfo {volume} {107}},\ \bibinfo {pages} {032412} (\bibinfo {year} {2023})}\BibitemShut {NoStop}%
\bibitem [{\citenamefont {Fukui}\ \emph {et~al.}(2018)\citenamefont {Fukui}, \citenamefont {Tomita}, \citenamefont {Okamoto},\ and\ \citenamefont {Fujii}}]{Fukui(2017)SURFACECODE}%
  \BibitemOpen
  \bibfield  {author} {\bibinfo {author} {\bibfnamefont {K.}~\bibnamefont {Fukui}}, \bibinfo {author} {\bibfnamefont {A.}~\bibnamefont {Tomita}}, \bibinfo {author} {\bibfnamefont {A.}~\bibnamefont {Okamoto}},\ and\ \bibinfo {author} {\bibfnamefont {K.}~\bibnamefont {Fujii}},\ }\bibfield  {title} {\bibinfo {title} {High-threshold fault-tolerant quantum computation with analog quantum error correction},\ }\href {https://doi.org/10.1103/PhysRevX.8.021054} {\bibfield  {journal} {\bibinfo  {journal} {Phys. Rev. X}\ }\textbf {\bibinfo {volume} {8}},\ \bibinfo {pages} {021054} (\bibinfo {year} {2018})}\BibitemShut {NoStop}%
\bibitem [{\citenamefont {Noh}\ and\ \citenamefont {Chamberland}(2020)}]{Noh(2019)SURFACE}%
  \BibitemOpen
  \bibfield  {author} {\bibinfo {author} {\bibfnamefont {K.}~\bibnamefont {Noh}}\ and\ \bibinfo {author} {\bibfnamefont {C.}~\bibnamefont {Chamberland}},\ }\bibfield  {title} {\bibinfo {title} {Fault-tolerant bosonic quantum error correction with the surface--gottesman-kitaev-preskill code},\ }\href {https://doi.org/10.1103/PhysRevA.101.012316} {\bibfield  {journal} {\bibinfo  {journal} {Phys. Rev. A}\ }\textbf {\bibinfo {volume} {101}},\ \bibinfo {pages} {012316} (\bibinfo {year} {2020})}\BibitemShut {NoStop}%
\bibitem [{\citenamefont {Tzitrin}\ \emph {et~al.}(2021)\citenamefont {Tzitrin}, \citenamefont {Matsuura}, \citenamefont {Alexander}, \citenamefont {Dauphinais}, \citenamefont {Bourassa}, \citenamefont {Sabapathy}, \citenamefont {Menicucci},\ and\ \citenamefont {Dhand}}]{TzitrinPRXQuantum(2021)}%
  \BibitemOpen
  \bibfield  {author} {\bibinfo {author} {\bibfnamefont {I.}~\bibnamefont {Tzitrin}}, \bibinfo {author} {\bibfnamefont {T.}~\bibnamefont {Matsuura}}, \bibinfo {author} {\bibfnamefont {R.~N.}\ \bibnamefont {Alexander}}, \bibinfo {author} {\bibfnamefont {G.}~\bibnamefont {Dauphinais}}, \bibinfo {author} {\bibfnamefont {J.~E.}\ \bibnamefont {Bourassa}}, \bibinfo {author} {\bibfnamefont {K.~K.}\ \bibnamefont {Sabapathy}}, \bibinfo {author} {\bibfnamefont {N.~C.}\ \bibnamefont {Menicucci}},\ and\ \bibinfo {author} {\bibfnamefont {I.}~\bibnamefont {Dhand}},\ }\bibfield  {title} {\bibinfo {title} {Fault-tolerant quantum computation with static linear optics},\ }\href {https://doi.org/10.1103/PRXQuantum.2.040353} {\bibfield  {journal} {\bibinfo  {journal} {PRX Quantum}\ }\textbf {\bibinfo {volume} {2}},\ \bibinfo {pages} {040353} (\bibinfo {year} {2021})}\BibitemShut {NoStop}%
\bibitem [{\citenamefont {Fowler}\ \emph {et~al.}(2012)\citenamefont {Fowler}, \citenamefont {Mariantoni}, \citenamefont {Martinis},\ and\ \citenamefont {Cleland}}]{SurfacecodesPRA}%
  \BibitemOpen
  \bibfield  {author} {\bibinfo {author} {\bibfnamefont {A.~G.}\ \bibnamefont {Fowler}}, \bibinfo {author} {\bibfnamefont {M.}~\bibnamefont {Mariantoni}}, \bibinfo {author} {\bibfnamefont {J.~M.}\ \bibnamefont {Martinis}},\ and\ \bibinfo {author} {\bibfnamefont {A.~N.}\ \bibnamefont {Cleland}},\ }\bibfield  {title} {\bibinfo {title} {Surface codes: Towards practical large-scale quantum computation},\ }\href@noop {} {\bibfield  {journal} {\bibinfo  {journal} {Phys. Rev. A}\ }\textbf {\bibinfo {volume} {86}},\ \bibinfo {pages} {032324} (\bibinfo {year} {2012})}\BibitemShut {NoStop}%
\bibitem [{\citenamefont {H\"anggli}\ \emph {et~al.}(2020)\citenamefont {H\"anggli}, \citenamefont {Heinze},\ and\ \citenamefont {K\"onig}}]{biassurface}%
  \BibitemOpen
  \bibfield  {author} {\bibinfo {author} {\bibfnamefont {L.}~\bibnamefont {H\"anggli}}, \bibinfo {author} {\bibfnamefont {M.}~\bibnamefont {Heinze}},\ and\ \bibinfo {author} {\bibfnamefont {R.}~\bibnamefont {K\"onig}},\ }\bibfield  {title} {\bibinfo {title} {Enhanced noise resilience of the surface--gottesman-kitaev-preskill code via designed bias},\ }\href@noop {} {\bibfield  {journal} {\bibinfo  {journal} {Phys. Rev. A}\ }\textbf {\bibinfo {volume} {102}},\ \bibinfo {pages} {052408} (\bibinfo {year} {2020})}\BibitemShut {NoStop}%
\bibitem [{\citenamefont {Zhang}\ \emph {et~al.}(2023)\citenamefont {Zhang}, \citenamefont {Wu},\ and\ \citenamefont {Guo}}]{XZZXsurfacecode}%
  \BibitemOpen
  \bibfield  {author} {\bibinfo {author} {\bibfnamefont {J.}~\bibnamefont {Zhang}}, \bibinfo {author} {\bibfnamefont {Y.-C.}\ \bibnamefont {Wu}},\ and\ \bibinfo {author} {\bibfnamefont {G.-P.}\ \bibnamefont {Guo}},\ }\bibfield  {title} {\bibinfo {title} {Concatenation of the gottesman-kitaev-preskill code with the xzzx surface code},\ }\href@noop {} {\bibfield  {journal} {\bibinfo  {journal} {Phys. Rev. A}\ }\textbf {\bibinfo {volume} {107}},\ \bibinfo {pages} {062408} (\bibinfo {year} {2023})}\BibitemShut {NoStop}%
\bibitem [{\citenamefont {Alves}\ \emph {et~al.}(2018)\citenamefont {Alves}, \citenamefont {Barros}, \citenamefont {Tasca}, \citenamefont {Souza},\ and\ \citenamefont {Khoury}}]{rulesoftransmode}%
  \BibitemOpen
  \bibfield  {author} {\bibinfo {author} {\bibfnamefont {G.~B.}\ \bibnamefont {Alves}}, \bibinfo {author} {\bibfnamefont {R.~F.}\ \bibnamefont {Barros}}, \bibinfo {author} {\bibfnamefont {D.~S.}\ \bibnamefont {Tasca}}, \bibinfo {author} {\bibfnamefont {C.~E.~R.}\ \bibnamefont {Souza}},\ and\ \bibinfo {author} {\bibfnamefont {A.~Z.}\ \bibnamefont {Khoury}},\ }\bibfield  {title} {\bibinfo {title} {Conditions for optical parametric oscillation with a structured light pump},\ }\href {https://doi.org/10.1103/PhysRevA.98.063825} {\bibfield  {journal} {\bibinfo  {journal} {Phys. Rev. A}\ }\textbf {\bibinfo {volume} {98}},\ \bibinfo {pages} {063825} (\bibinfo {year} {2018})}\BibitemShut {NoStop}%
\bibitem [{\citenamefont {Liu}\ \emph {et~al.}(2016)\citenamefont {Liu}, \citenamefont {Guo}, \citenamefont {Cai}, \citenamefont {Zhang},\ and\ \citenamefont {Gao}}]{liu2016GHZstate}%
  \BibitemOpen
  \bibfield  {author} {\bibinfo {author} {\bibfnamefont {K.}~\bibnamefont {Liu}}, \bibinfo {author} {\bibfnamefont {J.}~\bibnamefont {Guo}}, \bibinfo {author} {\bibfnamefont {C.}~\bibnamefont {Cai}}, \bibinfo {author} {\bibfnamefont {J.}~\bibnamefont {Zhang}},\ and\ \bibinfo {author} {\bibfnamefont {J.}~\bibnamefont {Gao}},\ }\bibfield  {title} {\bibinfo {title} {Direct generation of spatial quadripartite continuous variable entanglement in an optical parametric oscillator},\ }\href@noop {} {\bibfield  {journal} {\bibinfo  {journal} {Opt. Letters}\ }\textbf {\bibinfo {volume} {41}},\ \bibinfo {pages} {5178} (\bibinfo {year} {2016})}\BibitemShut {NoStop}%
\bibitem [{\citenamefont {Menicucci}\ \emph {et~al.}(2011)\citenamefont {Menicucci}, \citenamefont {Flammia},\ and\ \citenamefont {van Loock}}]{Menicucci2011star}%
  \BibitemOpen
  \bibfield  {author} {\bibinfo {author} {\bibfnamefont {N.~C.}\ \bibnamefont {Menicucci}}, \bibinfo {author} {\bibfnamefont {S.~T.}\ \bibnamefont {Flammia}},\ and\ \bibinfo {author} {\bibfnamefont {P.}~\bibnamefont {van Loock}},\ }\bibfield  {title} {\bibinfo {title} {Graphical calculus for gaussian pure states},\ }\href {https://doi.org/10.1103/PhysRevA.83.042335} {\bibfield  {journal} {\bibinfo  {journal} {Phys. Rev. A}\ }\textbf {\bibinfo {volume} {83}},\ \bibinfo {pages} {042335} (\bibinfo {year} {2011})}\BibitemShut {NoStop}%
\bibitem [{\citenamefont {Grosse}\ \emph {et~al.}(2006)\citenamefont {Grosse}, \citenamefont {Bowen}, \citenamefont {McKenzie},\ and\ \citenamefont {Lam}}]{GrossePRL96HarmonicEntanglement}%
  \BibitemOpen
  \bibfield  {author} {\bibinfo {author} {\bibfnamefont {N.~B.}\ \bibnamefont {Grosse}}, \bibinfo {author} {\bibfnamefont {W.~P.}\ \bibnamefont {Bowen}}, \bibinfo {author} {\bibfnamefont {K.}~\bibnamefont {McKenzie}},\ and\ \bibinfo {author} {\bibfnamefont {P.~K.}\ \bibnamefont {Lam}},\ }\bibfield  {title} {\bibinfo {title} {Harmonic entanglement with second-order nonlinearity},\ }\href {https://doi.org/10.1103/PhysRevLett.96.063601} {\bibfield  {journal} {\bibinfo  {journal} {Phys. Rev. Lett.}\ }\textbf {\bibinfo {volume} {96}},\ \bibinfo {pages} {063601} (\bibinfo {year} {2006})}\BibitemShut {NoStop}%
\bibitem [{\citenamefont {Grosse}\ \emph {et~al.}(2008)\citenamefont {Grosse}, \citenamefont {Assad}, \citenamefont {Mehmet}, \citenamefont {Schnabel}, \citenamefont {Symul},\ and\ \citenamefont {Lam}}]{GrossePRLHarmonicEntanglement}%
  \BibitemOpen
  \bibfield  {author} {\bibinfo {author} {\bibfnamefont {N.~B.}\ \bibnamefont {Grosse}}, \bibinfo {author} {\bibfnamefont {S.}~\bibnamefont {Assad}}, \bibinfo {author} {\bibfnamefont {M.}~\bibnamefont {Mehmet}}, \bibinfo {author} {\bibfnamefont {R.}~\bibnamefont {Schnabel}}, \bibinfo {author} {\bibfnamefont {T.}~\bibnamefont {Symul}},\ and\ \bibinfo {author} {\bibfnamefont {P.~K.}\ \bibnamefont {Lam}},\ }\bibfield  {title} {\bibinfo {title} {Observation of entanglement between two light beams spanning an octave in optical frequency},\ }\href@noop {} {\bibfield  {journal} {\bibinfo  {journal} {Phys. Rev. Lett.}\ }\textbf {\bibinfo {volume} {100}},\ \bibinfo {pages} {243601} (\bibinfo {year} {2008})}\BibitemShut {NoStop}%
\bibitem [{\citenamefont {Guo}\ \emph {et~al.}(2012)\citenamefont {Guo}, \citenamefont {Zhao},\ and\ \citenamefont {Li}}]{guoAPL2012HarmonicEntanglement}%
  \BibitemOpen
  \bibfield  {author} {\bibinfo {author} {\bibfnamefont {X.}~\bibnamefont {Guo}}, \bibinfo {author} {\bibfnamefont {J.}~\bibnamefont {Zhao}},\ and\ \bibinfo {author} {\bibfnamefont {Y.}~\bibnamefont {Li}},\ }\bibfield  {title} {\bibinfo {title} {Robust generation of bright two-color entangled optical beams from a phase-insensitive optical parametric amplifier},\ }\href@noop {} {\bibfield  {journal} {\bibinfo  {journal} {Applied Physics Letters}\ }\textbf {\bibinfo {volume} {100}} (\bibinfo {year} {2012})}\BibitemShut {NoStop}%
\bibitem [{\citenamefont {Yanagimoto}\ \emph {et~al.}(2023)\citenamefont {Yanagimoto}, \citenamefont {Nehra}, \citenamefont {Hamerly}, \citenamefont {Ng}, \citenamefont {Marandi},\ and\ \citenamefont {Mabuchi}}]{YanagimotoGKP(2023)PRXQuantum}%
  \BibitemOpen
  \bibfield  {author} {\bibinfo {author} {\bibfnamefont {R.}~\bibnamefont {Yanagimoto}}, \bibinfo {author} {\bibfnamefont {R.}~\bibnamefont {Nehra}}, \bibinfo {author} {\bibfnamefont {R.}~\bibnamefont {Hamerly}}, \bibinfo {author} {\bibfnamefont {E.}~\bibnamefont {Ng}}, \bibinfo {author} {\bibfnamefont {A.}~\bibnamefont {Marandi}},\ and\ \bibinfo {author} {\bibfnamefont {H.}~\bibnamefont {Mabuchi}},\ }\bibfield  {title} {\bibinfo {title} {Quantum nondemolition measurements with optical parametric amplifiers for ultrafast universal quantum information processing},\ }\href {https://doi.org/10.1103/PRXQuantum.4.010333} {\bibfield  {journal} {\bibinfo  {journal} {PRX Quantum}\ }\textbf {\bibinfo {volume} {4}},\ \bibinfo {pages} {010333} (\bibinfo {year} {2023})}\BibitemShut {NoStop}%
\bibitem [{\citenamefont {Walshe}\ \emph {et~al.}(2021)\citenamefont {Walshe}, \citenamefont {Alexander}, \citenamefont {Menicucci},\ and\ \citenamefont {Baragiola}}]{BaragiolaStreamlined(2021)}%
  \BibitemOpen
  \bibfield  {author} {\bibinfo {author} {\bibfnamefont {B.~W.}\ \bibnamefont {Walshe}}, \bibinfo {author} {\bibfnamefont {R.~N.}\ \bibnamefont {Alexander}}, \bibinfo {author} {\bibfnamefont {N.~C.}\ \bibnamefont {Menicucci}},\ and\ \bibinfo {author} {\bibfnamefont {B.~Q.}\ \bibnamefont {Baragiola}},\ }\bibfield  {title} {\bibinfo {title} {Streamlined quantum computing with macronode cluster states},\ }\href {https://doi.org/10.1103/PhysRevA.104.062427} {\bibfield  {journal} {\bibinfo  {journal} {Phys. Rev. A}\ }\textbf {\bibinfo {volume} {104}},\ \bibinfo {pages} {062427} (\bibinfo {year} {2021})}\BibitemShut {NoStop}%
\bibitem [{\citenamefont {Walshe}\ \emph {et~al.}(2023)\citenamefont {Walshe}, \citenamefont {Alexander}, \citenamefont {Matsuura}, \citenamefont {Baragiola},\ and\ \citenamefont {Menicucci}}]{Equivalentcluster}%
  \BibitemOpen
  \bibfield  {author} {\bibinfo {author} {\bibfnamefont {B.~W.}\ \bibnamefont {Walshe}}, \bibinfo {author} {\bibfnamefont {R.~N.}\ \bibnamefont {Alexander}}, \bibinfo {author} {\bibfnamefont {T.}~\bibnamefont {Matsuura}}, \bibinfo {author} {\bibfnamefont {B.~Q.}\ \bibnamefont {Baragiola}},\ and\ \bibinfo {author} {\bibfnamefont {N.~C.}\ \bibnamefont {Menicucci}},\ }\bibfield  {title} {\bibinfo {title} {Equivalent noise properties of scalable continuous-variable cluster states},\ }\href@noop {} {\bibfield  {journal} {\bibinfo  {journal} {Phys. Rev. A}\ }\textbf {\bibinfo {volume} {108}},\ \bibinfo {pages} {042602} (\bibinfo {year} {2023})}\BibitemShut {NoStop}%
\bibitem [{\citenamefont {Walshe}\ \emph {et~al.}(2020)\citenamefont {Walshe}, \citenamefont {Baragiola}, \citenamefont {Alexander},\ and\ \citenamefont {Menicucci}}]{teleportationQEC}%
  \BibitemOpen
  \bibfield  {author} {\bibinfo {author} {\bibfnamefont {B.~W.}\ \bibnamefont {Walshe}}, \bibinfo {author} {\bibfnamefont {B.~Q.}\ \bibnamefont {Baragiola}}, \bibinfo {author} {\bibfnamefont {R.~N.}\ \bibnamefont {Alexander}},\ and\ \bibinfo {author} {\bibfnamefont {N.~C.}\ \bibnamefont {Menicucci}},\ }\bibfield  {title} {\bibinfo {title} {Continuous-variable gate teleportation and bosonic-code error correction},\ }\href@noop {} {\bibfield  {journal} {\bibinfo  {journal} {Phys. Rev. A}\ }\textbf {\bibinfo {volume} {102}},\ \bibinfo {pages} {062411} (\bibinfo {year} {2020})}\BibitemShut {NoStop}%
\bibitem [{\citenamefont {Gidney}(2021)}]{Gidney2021stim}%
  \BibitemOpen
  \bibfield  {author} {\bibinfo {author} {\bibfnamefont {C.}~\bibnamefont {Gidney}},\ }\bibfield  {title} {\bibinfo {title} {Stim: a fast stabilizer circuit simulator},\ }\href {https://doi.org/10.22331/q-2021-07-06-497} {\bibfield  {journal} {\bibinfo  {journal} {{Quantum}}\ }\textbf {\bibinfo {volume} {5}},\ \bibinfo {pages} {497} (\bibinfo {year} {2021})}\BibitemShut {NoStop}%
\bibitem [{\citenamefont {Higgott}(2022)}]{higgott2022pymatching}%
  \BibitemOpen
  \bibfield  {author} {\bibinfo {author} {\bibfnamefont {O.}~\bibnamefont {Higgott}},\ }\bibfield  {title} {\bibinfo {title} {Pymatching: A python package for decoding quantum codes with minimum-weight perfect matching},\ }\href@noop {} {\bibfield  {journal} {\bibinfo  {journal} {ACM Transactions on Quantum Computing}\ }\textbf {\bibinfo {volume} {3}},\ \bibinfo {pages} {1} (\bibinfo {year} {2022})}\BibitemShut {NoStop}%
\bibitem [{\citenamefont {Fukui}\ \emph {et~al.}(2017)\citenamefont {Fukui}, \citenamefont {Tomita},\ and\ \citenamefont {Okamoto}}]{AnalogQuantumErrorPhysRevLett}%
  \BibitemOpen
  \bibfield  {author} {\bibinfo {author} {\bibfnamefont {K.}~\bibnamefont {Fukui}}, \bibinfo {author} {\bibfnamefont {A.}~\bibnamefont {Tomita}},\ and\ \bibinfo {author} {\bibfnamefont {A.}~\bibnamefont {Okamoto}},\ }\bibfield  {title} {\bibinfo {title} {Analog quantum error correction with encoding a qubit into an oscillator},\ }\href@noop {} {\bibfield  {journal} {\bibinfo  {journal} {Phys. Rev. Lett.}\ }\textbf {\bibinfo {volume} {119}},\ \bibinfo {pages} {180507} (\bibinfo {year} {2017})}\BibitemShut {NoStop}%
\end{thebibliography}%

\end{document}